\documentclass[final,5p,times,twocolumn]{elsarticle}

\usepackage{amsmath,amssymb,amsfonts}
\usepackage{balance}
\usepackage{graphicx}
\usepackage{url}
\usepackage{textcomp}
\usepackage{xcolor}
\usepackage{booktabs}
\usepackage{comment}
\usepackage{float}
\usepackage{array}
\usepackage{algorithm,algorithmic}
\usepackage{longtable}
\usepackage{tabularx}
\usepackage{tikz}
\usepackage{pgfplots}
\usepackage{microtype}
\usepackage[breaklinks]{hyperref}
\usepackage{pifont}
\usepackage{amsmath}
\usepackage{bm}

\newlength\savedindent

\journal{Future Generation Computer Systems}

\begin{document}

\begin{frontmatter}

\title{A Cross-Layered Multi-Drone Coordination for Medical Supply Delivery \\ during Disaster Response Management}

\author[mu]{Aneesh Calyam}
\ead{ackfw@missouri.edu}
\author[mu]{Subrahmanya Chandra Bhamidipati}
\ead{sb5q6@missouri.edu}
\author[mu]{Zack Murry}
\ead{zjmfrr@missouri.edu}
\author[mu]{Sharan Srinivas\corref{cor1}}
\ead{srinivassh@missouri.edu}
\ead[url]{https://engineering.missouri.edu/faculty/sharan-srinivas/}
\cortext[cor1]{Corresponding author}
\address[mu]{University of Missouri--Columbia, Columbia, MO 65211, USA}

\begin{abstract}
    Autonomous drone fleets have immense potential in medical supply delivery during disaster incident response. However, coordinating multiple drones in such settings introduces compounding challenges: dynamic environmental hazards such as wind, obstacles, and intermittent network connectivity, constrained energy budgets, and the need to serve patient locations fairly under strict deadlines and triage-based priority while optimizing schedule utilization. In this paper, we present CEDA, a novel Centralized Training with Decentralized Execution (CTDE) Double Deep Q-Network algorithm augmented with a QMIX-style monotonic mixing network for cooperative multi-drone medical delivery, designed to jointly optimize triage-priority-aware routing, multi-agent coordination, and energy-efficient navigation under dynamic uncertainty. CEDA introduces a Priority-Preserving Fair Scheduling strategy, in which a carefully structured reward function encodes triage priority weights, outcome-level fairness and priority-ordering rewards, and a set of complementary shaping mechanisms that collectively discourage systematic starvation. We evaluate CEDA in a simulated grid environment with a fleet of five drones serving up to fifty patients, featuring dynamic hazard zones, stochastic action failures, and dynamically spawning patients across three triage priority levels, as well as in a PX4 software-in-the-loop (SITL) validation using five X500 quadrotors controlled via MAVSDK (MAVLink Software Development Kit) in offboard position mode. Simulation results demonstrate that CEDA delivers over 95\% of spawned patients, reduces obstacle collisions by over 95\% across training, and achieves a weighted triage efficiency exceeding 0.90 alongside a Jain's Fairness Index of approximately 1.0 across triage classes. Critically, CEDA preserves clinical priority ordering, with Critical patients consistently receiving the fastest response times, while achieving low and comparable mortality across all triage classes, confirming that priority-weighted routing does not condemn Stable or Urgent patients to neglect. PX4 SITL validation further shows that the learned policy retains its triage-aware delivery behavior under continuous simulated flight and delayed-action conditions.
\end{abstract}

\begin{keyword}
Drone trajectory planning \sep Dynamic decision-making \sep Reinforcement Learning \sep Cross-layer Coordination
\end{keyword}

\end{frontmatter}

\section{Introduction}
\label{sec:intro}

Autonomous drone fleets have emerged as a promising solution for medical supply delivery during Disaster Response Management (DRM), where rapid intervention is critical to saving lives. Drones can deliver essential medical resources such as defibrillators, vaccines, and emergency drugs significantly faster than traditional ground-based logistics~\cite{nyaaba2021}, and can operate in physically inaccessible or hazardous regions where infrastructure may be damaged or unavailable.

\begin{figure}
    \centering
    \vspace{-12mm}
    \includegraphics[width=1\linewidth]{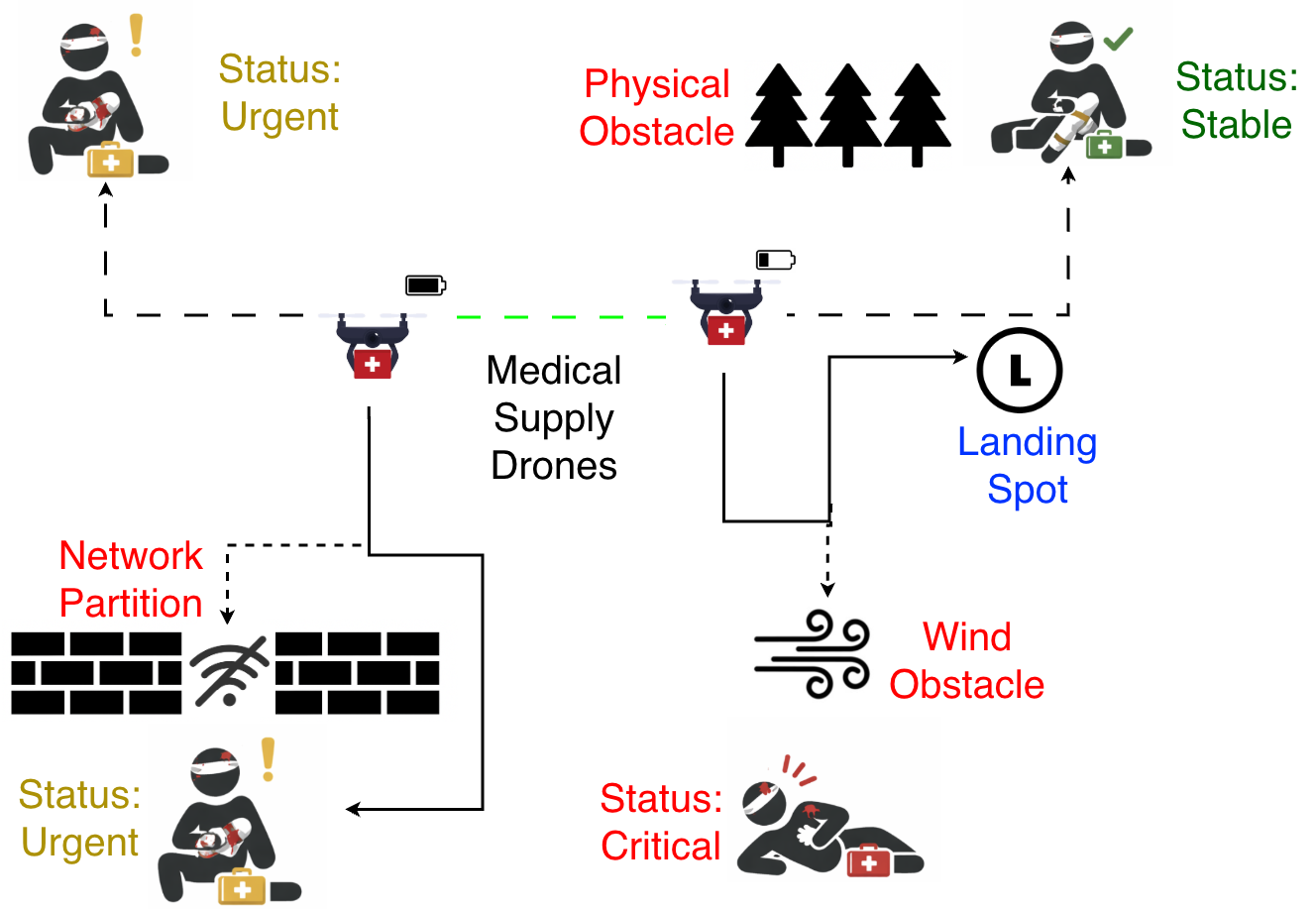}
    \caption{Illustration of multi-drone coordination in a dynamic DRM environment with triage-priority patients and energy constraints.}
    \label{fig:intro_fig}
\end{figure}

As illustrated in Figure~\ref{fig:intro_fig}, a multi-drone DRM system must simultaneously navigate physical hazards, maintain reliable communication under degraded network conditions, manage finite energy budgets, and serve multiple patients whose clinical urgency changes over time. This creates compounding interdependencies that conventional path planning and scheduling methods are not designed to handle jointly: a policy that ignores network reliability will fail when low-signal zones corrupt control commands, while a scheduler unaware of battery state will deplete drones before the mission is complete.

The challenge decomposes across three interdependent layers. At the \textit{drone path planning layer}, dynamic hazards including wind and obstacles require continuous adaptation under energy constraints. At the \textit{network awareness layer}, intermittent connectivity induces stochastic action failures, forcing routing decisions to internalize communication risk. At the \textit{application scheduling layer}, patients with heterogeneous clinical urgency --- Stable, Urgent, and Critical --- must be served under strict time constraints~\cite{aggarwal2023,abdelkarim2026hazard}, requiring triage-based prioritization balanced with equitable service. These layers cannot be resolved independently: the optimal path under energy constraints may be suboptimal under network reliability, and the clinically correct triage order may conflict with both.

Existing approaches address only subsets of these demands. Single-drone reinforcement learning frameworks navigate dynamic environments with wind, connectivity, and battery constraints~\cite{calyam2025dpp}, but assume a single agent and no clinical priority structure. While Earliest Deadline First (EDF)~\cite{liu1973scheduling} is theoretically optimal for deadline-constrained scheduling~\cite{buttazzo2011hard}, it treats all tasks as equal in importance, which is clinically inappropriate when Critical and Stable patients share similar timers, and priority inversion under continuous high-urgency arrivals causes starvation of lower-priority cases~\cite{sha1990priority,pinedo2016scheduling}, a well-documented failure mode requiring semantic scheduling that augments deadline ordering with clinical urgency weights.

Three key obstacles therefore remain in effective multi-drone DRM delivery:

\begin{itemize}
 \item \textbf{Cross-layer coordination under dynamic uncertainty:} Among the representative methods reviewed in Section~\ref{sec:relwork}, none jointly integrates routing under physical hazards, stochastic communication failures, and evolving patient urgency within a unified decision framework.

    \item \textbf{Triage-aware scheduling with fairness preservation:} A purely deadline-driven policy risks misallocating resources toward highest-weight patients while lower-priority patients expire unserved, requiring mechanisms beyond both EDF and weight-greedy strategies to remain simultaneously utilization-maximizing, deadline-aware, and equitable.

    \item \textbf{Heterogeneous patient deterioration under partial observability:} Each patient deteriorates at a different rate, yet agents observe only discrete triage weights and remaining timers rather than the underlying survival dynamics, requiring a survival model that creates consistent pressure toward timely service without relying on ground-truth deterioration parameters at execution time.
\end{itemize}

In this paper, we present a unified deep reinforcement learning framework for cooperative multi-drone DRM that addresses all these layers simultaneously. The framework makes the following contributions:

\begin{itemize}
    \item \textbf{CEDA Framework:} A novel CTDE Double Deep Q-Network augmented with a QMIX-style monotonic mixing network that jointly optimizes triage-priority-aware routing, multi-agent coordination, and energy-efficient navigation. The centralized learner leverages global state information during training, while each drone selects its own action independently at execution time using a synchronized shared task-state representation, so coordination is achieved through shared task-state information rather than peer-to-peer negotiation or explicit action-coordination messages.

    \item \textbf{Priority-Preserving Fair Scheduling:} A strategy that encodes triage priority weights and complementary fairness mechanisms directly into the DQN reward function, so equitable service behavior emerges across all patient classes rather than being imposed as an explicit constraint.

    \item \textbf{Per-Patient Survival Dynamics:} Each patient's clinical deterioration is modeled via an individual logistic survival curve with randomized decay parameters and escalation thresholds, producing time-varying triage weights that counteract static priority exploitation.

    \item \textbf{Multi-Scenario Evaluation and Baseline Comparison:} We evaluate the framework at an expanded five-drone, fifty-patient scale, benchmarking against contemporary MARL baselines (VDN, QMIX, MADDPG, MAPPO) and heuristic dispatch baselines (Nearest-Path, Priority-Path, Random) under identical environment, reward, and evaluation protocols, including in PX4 software-in-the-loop validation.

    \item \textbf{PX4 SITL Validation:} We validate the framework in PX4 Software-In-The-Loop (SITL) simulation, executing the real PX4 autopilot stack against a Gazebo physics simulator, demonstrating that the learned policy maintains its delivery and triage advantages over heuristic baselines through a realistic autopilot stack with takeoff, waypoint tracking, and landing transitions.
\end{itemize}

The remainder of this paper is organized as follows. Section~\ref{sec:relwork} reviews related work. Section~\ref{sec:problem} formulates the multi-drone medical delivery problem. Section~\ref{sec:algorithm} presents the proposed CEDA framework. Section~\ref{sec:results} provides experimental evaluation, and Section~\ref{sec:px4} details the PX4-SITL validation. Section~\ref{sec:conclusion} concludes the paper.
\section{Related Work}
\label{sec:relwork}


\newcommand{\cmarkc}{\ding{51}}
\newcommand{\xmarkc}{\ding{55}}
\newcommand{\tmarkc}{\ensuremath{\triangle}}
\newcommand{\cmarkH}{\ding{51}}

\begin{table*}[!t]
\centering
\caption{Feature comparison of representative UAV coordination, scheduling,
and navigation frameworks against CEDA. A feature is marked present only when
it is explicitly represented in the problem formulation, state, constraints,
objective, or reward.}
\label{tab:ceda-framework-comparison}

\small
\renewcommand{\arraystretch}{1.12}
\setlength{\tabcolsep}{4pt}

\begin{tabular*}{\textwidth}{
    @{\extracolsep{\fill}}
    >{\raggedright\arraybackslash}p{0.25\textwidth}
    *{8}{>{\centering\arraybackslash}p{0.065\textwidth}}
    @{}
}
\toprule
&
\multicolumn{1}{c}{Coordination}
&
\multicolumn{3}{c}{Scheduling}
&
\multicolumn{2}{c}{Clinical Service}
&
\multicolumn{2}{c}{Navigation}
\\

\cmidrule(lr){2-2}
\cmidrule(lr){3-5}
\cmidrule(lr){6-7}
\cmidrule(lr){8-9}

\textbf{Method}
& \textbf{MA}
& \textbf{TR}
& \textbf{DL}
& \textbf{DR}
& \textbf{SV}
& \textbf{FR}
& \textbf{EN}
& \textbf{HN}
\\

\midrule

Bhamidipati et al. \cite{bhamidipati2025q}
& \xmarkc
& \xmarkc
& \xmarkc
& \xmarkc
& \xmarkc
& \xmarkc
& \xmarkc
& \cmarkc
\\

Calyam et al. \cite{calyam2025dpp}
& \xmarkc
& \xmarkc
& \xmarkc
& \xmarkc
& \xmarkc
& \xmarkc
& \cmarkc
& \cmarkc
\\

Chen et al. \cite{chen2025cooperative}
& \cmarkc
& \xmarkc
& \cmarkc
& \xmarkc
& \xmarkc
& \xmarkc
& \xmarkc
& \xmarkc
\\

Abdel Karim et al. \cite{abdelkarim2026hazard}
& \cmarkc
& \xmarkc
& \xmarkc
& \xmarkc
& \xmarkc
& \xmarkc
& \cmarkc
& \cmarkc
\\

Guven and Parlak \cite{guven2026uavmarl}
& \cmarkc
& \cmarkc
& \cmarkc
& \cmarkc
& \xmarkc
& \xmarkc
& \tmarkc
& \tmarkc
\\

Kong et al. \cite{kong2024multi}
& \cmarkc
& \xmarkc
& \xmarkc
& \xmarkc
& \xmarkc
& \xmarkc
& \xmarkc
& \cmarkc
\\

Peng et al. \cite{peng2025multi}
& \cmarkc
& \xmarkc
& \xmarkc
& \cmarkc
& \xmarkc
& \xmarkc
& \cmarkc
& \xmarkc
\\

Zerrouk et al. \cite{zerrouk2025increasing}
& \cmarkc
& \xmarkc
& \xmarkc
& \xmarkc
& \xmarkc
& \xmarkc
& \tmarkc
& \tmarkc
\\

\midrule

\textbf{CEDA}
& \cmarkH
& \cmarkH
& \cmarkH
& \cmarkH
& \cmarkH
& \cmarkH
& \cmarkH
& \cmarkH
\\

\bottomrule
\end{tabular*}

\vspace{1.5mm}

\begin{minipage}{\textwidth}
\footnotesize
\raggedright

\cmarkc\ Present
\qquad
\xmarkc\ Absent
\qquad
\tmarkc\ Partial or narrower implementation.

\vspace{0.8mm}

\textbf{MA}: explicit multi-agent or multi-vehicle coordination;
\textbf{TR}: clinical triage or urgency classes;
\textbf{DL}: explicit per-request delivery deadlines;
\textbf{DR}: requests that arrive during execution;
\textbf{SV}: individualized patient survival or deterioration model;
\textbf{FR}: explicit fairness objective, constraint, metric, or starvation control;
\textbf{EN}: battery/endurance represented in routing or navigation decisions;
\textbf{HN}: environmental- or network-hazard information represented in routing
or navigation decisions.

\vspace{0.8mm}

\textit{Partial codings:}
Guven and Parlak impose a coarse flight-time/endurance limit and range-limited
communication, but do not use a detailed battery state or explicit channel/environmental
hazard process. Zerrouk et al. use vehicle-dependent coverage capability and static
no-fly-zone avoidance, but do not perform online battery- or hazard-adaptive navigation.
\end{minipage}

\vspace{-3mm}
\end{table*}

This section surveys four thematic areas relevant to the problem studied: drone-based disaster response and medical supply delivery, reinforcement learning for UAV navigation with cross-layer awareness, real-time scheduling under deadline and priority constraints, and fairness-aware multi-agent task assignment. 

Table \ref{tab:ceda-framework-comparison} presents a dedicated comparison between CEDA and eight other state-the-art UAV coordination, scheduling, and navigation methods. The comparison uses eight features covering multi-agent coordination, triage, deadlines, dynamic requests, survival modeling, fairness, energy awareness, and hazard or network awareness.

\subsection{Drone-Based Medical and Emergency Delivery}

The use of UAVs for emergency medical supply delivery has attracted sustained research interest in environments where ground infrastructure is damaged or overwhelmed. Operational challenges of drone-based medical logistics in low-resource settings have been surveyed extensively, with regulatory, logistical, and technical barriers identified as the primary obstacles~\cite{nyaaba2021,sharma2024}. Field studies have reported that drones frequently lose telemetry mid-flight due to poor connectivity and that adverse weather disrupts pre-planned routes unpredictably~\cite{aggarwal2023}, directly motivating the network-aware and hazard-aware design of the present work.

DQN-based path planning has been applied to urban emergency resource scheduling with priority-ordered demand points and constrained routing~\cite{zhao2023dqn}, while cooperative deep reinforcement learning has been proposed for jointly scheduling heterogeneous UAVs in emergency medical delivery with soft deadlines~\cite{chen2025cooperative}. Hazard-aware multi-UAV frameworks integrating DQN-based routing with weather, battery, and hazard severity in the state representation~\cite{abdelkarim2026hazard} share several design principles with ours, but do not model triage urgency, deadline-driven prioritization, or fairness across service classes. The closest existing work in terms of clinical urgency modeling uses PPO-based MARL with three urgency classes across fleets of up to 20 UAVs~\cite{guven2026uavmarl}, but models energy and hazards only narrowly through flight-time and communication-range limits. It does not include detailed battery depletion, explicit environmental-hazard processes, or explicit fairness mechanisms. Table~\ref{tab:ceda-framework-comparison} summarizes these differences directly. Chen et al. ~\cite{chen2025cooperative} represent multi-agent coordination and delivery deadlines; Abdel Karim et al. ~\cite{abdelkarim2026hazard} represent multi-agent coordination together with energy and hazard-aware navigation; and Guven and Parlak et al.  ~\cite{guven2026uavmarl} represent multi-agent coordination, triage, deadlines, and dynamically arriving requests, with narrower energy and hazard representations. In comparison, CEDA represents all eight features, including the survival-modeling and fairness features absent from these three frameworks.

\subsection{Reinforcement Learning and Cross-Layer Awareness in UAV Path Planning}

Conventional UAV routing strategies treat mobility and network connectivity as independent concerns, limiting situational awareness under simultaneous environmental and network disruptions. Tabular Q-learning has been applied to small discretized grid environments due to its interpretability~\cite{calyam2025dpp}, but does not scale to richer state representations~\cite{zhou2023ddqn, bhamidipati2025q}. Hierarchical reinforcement learning decomposes navigation into subtasks for improved robustness~\cite{liu2023hierarchical}, while memory-based DRL leverages recurrent models to handle partial observability~\cite{singla2019memory}. Despite these advances, existing methods do not integrate cross-layer decision-making that unifies mobility and connectivity awareness into a single policy.

Cross-layer design has emerged as a principled approach to embedding situational awareness into routing decisions, propagating information from physical-layer signal quality up to application-layer task urgency~\cite{abdelkarim2026hazard,zhang2019cellular}. DQN and its Double DQN variant achieve faster convergence than tabular Q-learning in obstacle-rich UAV environments~\cite{zhou2023ddqn}, and experience replay with target network stabilization provide the training stability required in multi-agent settings where simultaneous agent learning introduces non-stationarity~\cite{mnih2015human}. For multi-drone settings, CTDE has been identified as particularly effective for complex partially observable environments~\cite{survey2025marl,kong2024multi}, owing to the centralized critic's ability to condition on joint observations during training while execution remains fully decentralized. The navigation columns of Table~\ref{tab:ceda-framework-comparison} place these contributions in context. Bhamidipati et al.~\cite{bhamidipati2025q} represent hazard-aware navigation, Calyam et al.~\cite{calyam2025dpp} represent both energy and hazard-aware navigation, and Kong et al.~\cite{kong2024multi} combine multi-agent coordination with hazard-aware navigation. However, the corresponding scheduling and clinical-service columns show that these capabilities are not combined with triage, dynamic requests, individualized survival modeling, and fairness as they are in CEDA.

Among the studies reviewed here, none incorporates cross-layer state representations that jointly encode physical, network, and application-layer features, nor do they embed deadline-driven task prioritization or fairness constraints into their reward structures. The present work addresses all three gaps simultaneously.

\subsection{Real-Time Scheduling, Fairness, and Priority Constraints}

Earliest Deadline First scheduling, proven optimal for preemptive single-processor scheduling~\cite{liu1973scheduling}, guarantees a feasible schedule whenever one exists and has been extended to distributed and networked systems~\cite{buttazzo2011hard}. Its application to network measurement scheduling demonstrated significant improvements in schedulable utilization over round-robin and token-passing approaches~\cite{calyam2007ieeetc}. Pure EDF, however, treats all tasks as equal in importance, leading to clinically suboptimal outcomes in triage scenarios where urgency must supersede strict deadline ordering. Priority inversion, a well-documented failure mode of fixed-priority schedulers~\cite{sha1990priority,buttazzo2011hard}, is addressed by semantic priority scheduling, which augments deadline-based ordering with priority weights reflecting task importance, consistently outperforming pure EDF on satisfaction ratio and average stretch metrics~\cite{calyam2007ieeetc,calyam2014jnsm}.

A persistent companion challenge is the starvation of low-priority tasks when high-priority tasks arrive continuously. The Jain's Fairness Index~\cite{jain1984fairness} formalizes equity of resource distribution and has been widely adopted alongside throughput-focused metrics in network resource allocation~\cite{calyam2014jnsm}. In the drone delivery literature, fairness has received comparatively little attention. Existing multi-drone systems optimize aggregate objectives without accounting for whether all patients across triage levels receive service within acceptable time horizons~\cite{peng2025multi,zerrouk2025increasing}.This distinction is visible in Table~\ref{tab:ceda-framework-comparison}. Peng et al.~\cite{peng2025multi} represent multi-agent coordination, dynamically arriving requests, and energy-aware routing, whereas Zerrouk et al.~\cite{zerrouk2025increasing} represent multi-agent coordination with narrower energy and hazard capabilities. Neither framework explicitly represents clinical triage, individualized survival dynamics, or fairness, all of which are incorporated by CEDA.

The present paper addresses both challenges jointly by embedding EDF-based semantic priority scheduling directly into the DQN reward function, augmented by a dynamic weight escalation mechanism in which patient weights increase over time according to individual logistic survival curves. This time-varying priority signal counteracts priority inversion, while the triage efficiency metric $\eta=\sum_{i\in S}w_i/\sum_{i\in\zeta}w_i$ measures priority-weighted service completion. Further, the uniform per-step pending-patient penalty applied regardless of weight class discourages neglect of lower-priority patients. Fairness across triage classes is evaluated separately using Jain's Fairness Index and class-specific service outcomes in Section~\ref{subsec:triage}. 

Among the studies reviewed here, none combines EDF-based semantic priority scheduling, dynamic per-patient priority escalation, and fairness-aware reward mechanisms within a DRL framework for multi-drone medical delivery.

\section{Medical Supply Delivery Problem Formulation}
\label{sec:problem}

\begin{figure}[h]
    \centering
    \includegraphics[width=1\linewidth]{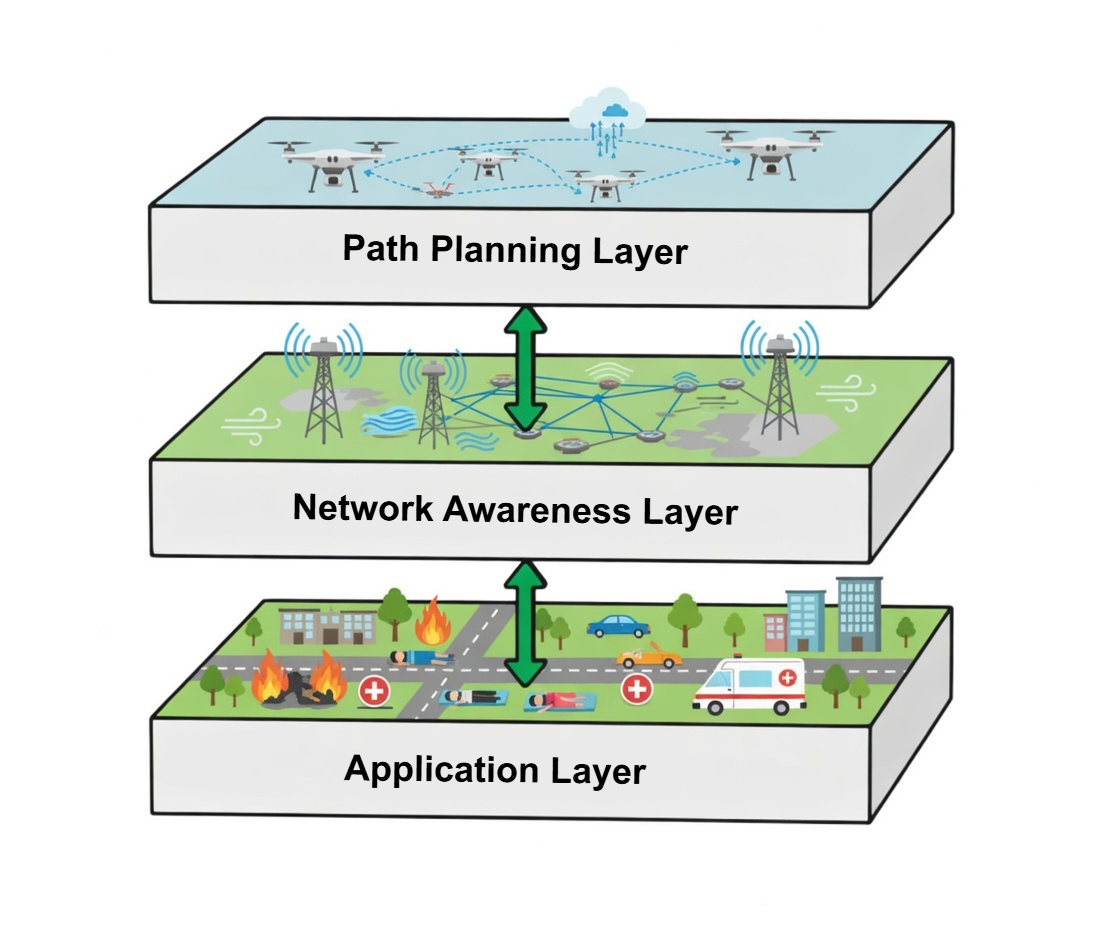}
    \vspace{-14mm}
    \caption{Cross-layer architecture governing drone routing decisions across the drone path planning, network awareness, and application scheduling layers.}
    \label{fig:problem_crosslayer}
\end{figure}

This section formalizes the multi-drone medical delivery scheduling problem under triage-constrained, deadline-driven conditions. The formulation defines the grid environment, the structure of delivery tasks, the priority model governing task urgency, and the dual objectives of utilization and triage-aware prioritization that jointly shape the scheduling policy. The cross-layer architecture governing drone routing decisions across these layers is illustrated in Figure~\ref{fig:problem_crosslayer}. The main symbols and their meanings are shown in Table~\ref{tab:notation}.

\begin{table}[h]
\centering
\caption{Notations}
\label{tab:notation}
\small
\begin{tabular}{@{}ll@{}}
\toprule
\textbf{Notation} & \textbf{Description} \\
\midrule
$\mathcal{G} = (\mathcal{V}, \mathcal{E})$ & Discrete grid-based environment \\
$\mathcal{O}$ & Set of static obstacle cells; $\mathcal{O} \subset \mathcal{V}$ \\
$\mathcal{D} = \{d_1, \ldots, d_N\}$ & Fleet of $N$ autonomous delivery drones \\
$\mathcal{C} = \{c_1, \ldots, c_M\}$ & Set of $M$ injured patients \\
$z_k$ & Dedicated landing zone of drone $d_k$ \\
$\tau_i = (\ell_i, \rho_i, w_i, d_i, e_i)$ & Real-time scheduling task for patient $c_i$ \\
$\rho_i$ & Triage level; $\rho_i \in \{\text{Critical, Urgent, Stable}\}$ \\
$\mathcal{W} = \{w_{\text{std}}, w_{\text{urg}}, w_{\text{crit}}\}$ & Priority weights; $w_{\text{crit}} > w_{\text{urg}} > w_{\text{std}} > 0$ \\
$\zeta = \{\tau_1, \ldots, \tau_M\}$ & Complete task set \\
$\mathcal{S} \subseteq \zeta$ & Tasks completed before their deadlines \\
$U$ & Delivery utilization; $U = |\mathcal{S}| / |\zeta|$ \\
$T_{\text{ep}}$ & Episode horizon (environment time steps) \\
$T_{\max}$ & Timer-normalization constant \\
$T_{\text{init}}$ & Countdown timer initialization value per patient; $d_i = T_{\text{init}}\ \forall i$ \\
$t_i^{\text{rem}}$ & Remaining timer of patient $c_i$ at delivery \\
$R_i$ & Timer- and triage-scaled delivery reward \\
$r_{\text{close}}$ & Manhattan distance threshold for penalty \\
$\Delta_{\text{base}},\ \Delta_{\text{wind}}$ & Baseline and wind-zone battery drain rates \\
$\eta$ & Weighted triage efficiency\\
$S_i(t)$ & Survival probability of patient $c_i$ at time $t$ \\
$a_i,\ b_i$ & Steepness and centering of logistic curve \\
$\theta_i^{\text{serious}},\ \theta_i^{\text{critical}}$ & Survival thresholds for triage escalation \\
$w_i(t)$ & Time-varying triage weight of patient $c_i$ \\
$\alpha$ & Trade-off scalar between $U$ and $\eta$ \\
$p_{\text{fail}}$ & Command failure prob. in low-signal zones \\
$b^i_t$ & Remaining battery of drone $d_i$ at time step $t$ \\
$\mathcal{S}_t = [o^1_t, \ldots, o^N_t]$ & Joint state observed at time step $t$ \\
$\mathcal{A}_t$ & Actions; $\{\textsc{north, south, west, east, hover, land}\}$ \\
\bottomrule
\normalsize
\end{tabular}
\end{table}

\subsection{Environment Model}

We consider a discrete grid-based environment $\mathcal{G} = (\mathcal{V}, \mathcal{E})$, where $\mathcal{V}$ denotes the set of navigable grid cells and $\mathcal{E}$ represents the set of edges connecting adjacent cells. Certain cells $\mathcal{O} \subset \mathcal{V}$ are designated as static obstacles and are impassable to all drones. A fleet of $N$ autonomous delivery drones $\mathcal{D} = \{d_1, d_2, \ldots, d_N\}$ operates within $\mathcal{G}$, each capable of carrying a medical payload sufficient to serve any patient in the environment. A set of $M$ injured patients $\mathcal{C} = \{c_1, c_2, \ldots, c_M\}$ is distributed across the grid, with each patient $c_j$ located at a known cell $\ell_j \in \mathcal{V}$ and requiring a time-sensitive medical delivery. Each drone $d_k$ is assigned a dedicated landing zone $z_k \in \mathcal{V}$ to which it must return upon completing all assigned deliveries. The formulation generalizes to arbitrary fleet and patient-pool sizes; concrete values of $N$ and $M$ used in experiments are reported in Section~\ref{sec:setup}.

Each medical delivery requirement is modeled as a real-time scheduling task $\tau_i$, defined formally as:

\begin{equation}
    \tau_i = \left( \ell_i,\ \rho_i,\ w_i,\ d_i,\ e_i \right)
\end{equation}

where $\ell_i \in \mathcal{V}$ is the grid location of patient $c_i$; $\rho_i \in \{\text{Critical},\ \text{Urgent},\ \text{Stable}\}$ is the assigned triage level; $w_i \in \mathcal{W}$ is the integer priority weight derived from $\rho_i$ (defined in Section~\ref{sec:triage}); $d_i = T_{\text{init}}$ is the delivery deadline, a fixed countdown value shared uniformly across all patients regardless of triage level (Section~\ref{sec:fairness}); and $e_i$ is the estimated execution time for a drone to navigate to $\ell_i$ from its current position.

This formulation mirrors the real-time task model of~\cite{calyam2007ieeetc}. The patient pool is not fixed at the start of each episode; new patients spawn dynamically at regular intervals of $\Delta_{\text{spawn}}$ steps throughout the mission, continuously refreshing $\zeta$ and preventing the agents from converging on a static delivery route.

\subsection{Scheduling Objectives}

The scheduling problem studied requires jointly balancing utilization and priority-weighted service efficiency: maximizing the fraction of deliveries completed on time, while ensuring higher-priority patients are served proportionally faster. Fairness across triage classes is not a separate objective here; it is handled through the reward mechanisms of Section~\ref{sec:fairness} and evaluated empirically in Section~\ref{subsec:triage}.

\subsubsection{Delivery Utilization}

Let $\mathcal{S} \subseteq \zeta$ denote the subset of tasks completed before their respective deadlines. Delivery utilization is defined as:

\begin{equation}
    U = \frac{|\mathcal{S}|}{|\zeta|}
\end{equation}

A scheduler maximizing $U$ ensures that as many patients as possible receive medical deliveries within their required time windows. The foundational scheduling heuristic we adopt is \textit{Earliest Deadline First} (EDF)~\cite{calyam2007ieeetc}: provides the deadline-awareness principle underlying the scheduling formulation: tasks with less remaining time become increasingly costly to delay. Rather than imposing an explicit EDF dispatch rule, CEDA exposes remaining-time information to the policy and incorporates timer-dependent reward shaping, as described in Sections 3.2.3 and 4.1.4. This allows deadline pressure to interact with triage priority, hazard exposure, and energy state during multi-drone action selection. 

\subsubsection{Triage-Aware Priority Model}
\label{sec:triage}
In emergency medical scenarios, delivery tasks are not equal in urgency, and a scheduler that treats them uniformly risks misallocating drone resources at the cost of patient outcomes. We adopt a \textit{semantic priority} model~\cite{calyam2014jnsm} that assigns each task a priority weight $w_i \in \mathcal{W}$ reflecting its clinical importance relative to other tasks in $\zeta$, where $\mathcal{W} = \{w_{\text{std}},\ w_{\text{urg}},\ w_{\text{crit}}\}$ with $w_{\text{crit}} > w_{\text{urg}} > w_{\text{std}} > 0$. The weight is a function of the triage level $\rho_i$, based on the discrete triage priority levels used by medical professionals in emergency medical scenarios~\cite{christian2019triage}:

\begin{equation}
    w_i =
    \begin{cases}
        w_{\text{crit}} & \text{if } \rho_i = \text{Critical} \\
        w_{\text{urg}}  & \text{if } \rho_i = \text{Urgent}   \\
        w_{\text{std}}  & \text{if } \rho_i = \text{Stable}
    \end{cases}
\end{equation}

\noindent Concrete values of $\mathcal{W}$ are reported in Section~\ref{sec:setup}. Each successful delivery yields a reward scaled by $w_i$, biasing the learned policy toward clinically critical patients while preserving incentives to service all triage levels. The semantic weight $w_i$ acts as an urgency amplifier that allows clinical priority to supersede strict deadline ordering when warranted, analogous to the Priority and Deadline Scheduling (PDS) formulation described in~\cite{calyam2014jnsm}.

\subsubsection{Priority-Preserving Fair Scheduling}
\label{sec:fairness}
We augment the semantic priority model with a carefully constructed reward structure that introduces fairness as a systemic property of the scheduler. Rather than imposing a single explicit fairness constraint, we engineer multiple interacting mechanisms, spanning the discrete triage classes of Section~\ref{sec:triage}, the survival-based weight escalation model of Section~\ref{sec:survival}, and the reward and penalty terms described below, that collectively discourage systematic neglect of any triage level. The reward for successfully delivering to patient $c_i$ is:
\begin{equation}
    R_i = R_{\text{goal}} \cdot \frac{t_i^{\text{rem}}}{T_{\max}} \cdot w_i(t)
\end{equation}
where $t_i^{\mathrm{rem}}$ is the patient's remaining timer at delivery, and $T_{\max}$ denotes the fixed timer-normalization constant used in the state and reward formulations. The patient countdown duration and episode horizon are specified separately in Section~5.1. The normalized ratio $t_i^{\text{rem}}/T_{\max}$ decreases as response time elapses, discouraging delayed service; it does not itself escalate priority. Clinical escalation is instead captured by the time-varying triage weight $w_i(t)$ (Section~\ref{sec:survival}), which transitions among Stable, Urgent, and Critical based on estimated survival probability. A per-step penalty for pending patients and a penalty for expiration without delivery further discourage prolonged delays, so the reward combines a continuous measure of remaining response time with discrete, clinically interpretable priority changes.

\paragraph{Uniform pending-patient penalty} At each time step, every pending patient contributes the same response-wait penalty, independent of triage weight. This class-independent penalty creates continuous pressure against prolonged neglect of lower-priority patients. If a patient's timer expires before delivery, the separate patient-death penalty in Eq.~(11) is applied with priority-dependent scaling.


\paragraph{Nearest-patient potential shaping} At every time step, each agent receives a potential-based shaping reward proportional to the reduction in Manhattan distance to its nearest undelivered patient, regardless of that patient's weight, continuously redistributing agent attention across the full patient pool.

\paragraph{Spatial separation via closeness penalty} Agents incur a penalty whenever the Manhattan distance between any pair of active drones falls below a threshold $r_{\text{close}}$:

\begin{equation}
    P_{\text{close}}^{ij} =
    \begin{cases}
        -C & \text{if } \|pos_{d_i} - pos_{d_j}\|_1 < r_{\text{close}} \\
        0  & \text{otherwise}
    \end{cases}
    \qquad \forall\, i \neq j
\end{equation}

\noindent with the fleet-level penalty given by the sum over all such pairs, $\sum_{i \neq j} P_{\text{close}}^{ij}$. This discourages co-location, implicitly partitioning the grid among active drones and reducing the probability that multiple drones converge on the same high-weight patients while others go unserved.

\paragraph{Dynamic patient spawning} New patients are activated at a fixed interval of $\Delta_{\text{spawn}}$ steps throughout each episode, preventing agents from learning a fixed greedy route optimized for the initial patient layout and making pure weight-greedy strategies sub-optimal over the full episode horizon.

\paragraph{Battery constraints} Each drone carries a finite battery depleting at a baseline rate $\Delta_{\text{base}}$ per step, with an elevated rate $\Delta_{\text{wind}}$ inside wind zones. The low-battery and depletion penalties discourage travel strategies that concentrate effort on a small subset of the patient pool, reinforcing broad coverage.

\medskip
\noindent The overall effect of these six mechanisms is a set of \textit{complementary fairness mechanisms}: no single component guarantees equitable service, but their joint interaction, together with the explicit fairness and triage-ordering outcome rewards of Section~\ref{subsec:reward_design}, creates consistent pressure toward serving patients of all triage levels within their deadlines. We quantify this empirically in Section~\ref{sec:results} via per-weight delivery counts and a weighted triage efficiency score:

\begin{equation}
    \eta = \frac{\sum_{i \in \mathcal{S}} w_i}{\sum_{i \in \zeta} w_i}
\end{equation}

\noindent where $\eta = 1$ indicates that all spawned patients were delivered, weighted by clinical priority.

\subsection{Patient Survival Dynamics and Triage Escalation}
\label{sec:survival}

The triage weight $w_i$ assigned to patient $c_i$ at spawn time is not static. In realistic emergency scenarios, a patient's clinical condition deteriorates over time, and a delivery adequate for a stable patient becomes critical if sufficiently delayed. We model each patient's survival probability using an individual logistic decay function:

\begin{equation}
\label{eq:survival}
S_i(t) = \frac{1}{1 + \exp\!\left(a_i \cdot t_i^{\text{elapsed}} - b_i\right)}
\end{equation}

\noindent where $t_i^{\text{elapsed}}$ is the number of time steps since activation, $a_i > 0$ controls steepness of decline, and $b_i > 0$ controls when that decline is centered. Each patient receives individually sampled values of $a_i$ and $b_i$ conditioned on their initial triage level $\rho_i$, as summarized in Table~\ref{tab:decay_params}, producing heterogeneous deterioration trajectories. Concrete parameter ranges are reported in Section~\ref{sec:setup}.

\begin{table}[h]
\centering
\caption{Per-patient decay parameter ranges by initial triage level}
\label{tab:decay_params}
\small
\begin{tabular}{@{}p{1.4cm}ccp{1.6cm}@{}}
\toprule
\textbf{Level} $\rho_i$ & $a_i$ range & $b_i$ range & \textbf{Escalation} \\
\midrule
Stable   & $[a_{\text{std}}^-,\ a_{\text{std}}^+]$ & $[b_{\text{std}}^-,\ b_{\text{std}}^+]$ & Slow, late \\
Urgent   & $[a_{\text{urg}}^-,\ a_{\text{urg}}^+]$ & $[b_{\text{urg}}^-,\ b_{\text{urg}}^+]$ & Moderate \\
Critical & $[a_{\text{crit}}^-,\ a_{\text{crit}}^+]$ & $[b_{\text{crit}}^-,\ b_{\text{crit}}^+]$ & Steep, early \\
\bottomrule
\end{tabular}
\normalsize
\end{table}

The discrete triage weight $w_i(t) \in \mathcal{W}$ is derived from $S_i(t)$ via two individually sampled thresholds $\theta_i^{\text{serious}}$ and $\theta_i^{\text{critical}}$:

\begin{equation}
\label{eq:triage_escalation}
w_i(t) =
\begin{cases}
w_{\text{crit}} & \text{if } S_i(t) < \theta_i^{\text{critical}} \\
w_{\text{urg}}  & \text{if } \theta_i^{\text{critical}} \leq S_i(t) < \theta_i^{\text{serious}} \\
w_{\text{std}}  & \text{otherwise}
\end{cases}
\end{equation}

\noindent subject to $\theta_i^{\text{critical}} < \theta_i^{\text{serious}} - \delta_{\theta}$, where $\delta_{\theta}$ is a minimum separation margin, producing a fully individualized deterioration profile $(a_i, b_i, \theta_i^{\text{serious}}, \theta_i^{\text{critical}})$ per patient.

Because $w_i(t)$ changes discretely at escalation thresholds while the timer ratio $t_i^{\text{rem}}/T_{\max}$ decreases continuously, the combined reward $R_i$ is not assumed to decrease monotonically across triage transitions; the class-specific response-time results in Section~\ref{sec:results} assess whether the learned policy nonetheless responds to clinical urgency without systematically delaying lower-priority patients. Decay parameters and escalation thresholds are not observable by the agents, so the policy relies only on the current triage weight and remaining timer under heterogeneous, partially observed deterioration dynamics.

\subsection{Formal Problem Statement}

The medical drone delivery scheduling problem studied is fundamentally a cross-layer optimization problem. Routing decisions made at the physical layer, choosing which cell to move to next, directly affect the network-layer reliability of subsequent commands and the application-layer reward obtainable from patient deliveries. These three layers cannot be decoupled: a policy that optimizes any single layer in isolation will be suboptimal or infeasible when the others impose binding constraints. We therefore formalize the problem as a unified cross-layer joint optimization.

Given a grid environment $\mathcal{G}$, a fleet $\mathcal{D}$ of $N$ drones, and a dynamically evolving task set $\zeta$ of up to $M$ medical delivery requests, each characterized by a patient location $\ell_i$, triage level $\rho_i$, time-varying semantic priority weight $w_i(t) \in \mathcal{W}$, deadline $d_i$, and execution time $e_i$, find a joint movement policy $\pi: \mathcal{S}_t \rightarrow \mathcal{A}_t$ for all drones in $\mathcal{D}$ that jointly maximizes delivery utilization $U$ and triage efficiency $\eta$:

\begin{equation}
\pi^* = \arg\max_{\pi}\ \alpha \cdot U(\pi) + (1 - \alpha) \cdot \eta(\pi)
\end{equation}

\noindent subject to the following cross-layer constraints:

\begin{itemize}
\item \textbf{Physical layer:} Battery constraint $b^i_t \geq 0\ \forall\ i, t$; obstacle avoidance $\text{pos}(d_i, t) \notin \mathcal{O}\ \forall\ i, t$; elevated energy drain $\Delta_{\text{wind}}$ inside wind zones; inter-agent collision avoidance $\text{pos}(d_a, t) \neq \text{pos}(d_b, t)\ \forall\ a \neq b$.

\item \textbf{Network layer:} Stochastic command failure with probability $p_{\text{fail}}$ inside low-signal zones, causing intended movement actions to be silently dropped; signal-degraded cells incur additional step penalties that modulate effective path cost.

\item \textbf{Application layer:} Each delivery must be completed before the patient's deadline $d_i$; triage weights $w_i(t)$ escalate over time according to the per-patient logistic survival model defined in Section~\ref{sec:survival}; patients that expire without service incur a shared penalty $P_{\text{death}}$ applied symmetrically to all agents.
\end{itemize}

\noindent The cross-layer coupling is what distinguishes this problem from classical vehicle routing or real-time scheduling. In a purely application-layer formulation, the policy would navigate directly to the highest-priority patient along the shortest path. In a purely physical-layer formulation, the policy would minimize energy expenditure by avoiding wind zones regardless of patient urgency. In a purely network-layer formulation, the policy would route around low-signal zones to maximize command reliability. The cross-layer formulation requires the policy to reason simultaneously over all three cost surfaces, since the optimal trajectory under one layer is generally infeasible or suboptimal under the others.

The joint state $\mathcal{S}_t = [o^1_t,\ \ldots,\ o^N_t]$ observed at each time step encodes features from all three layers: patient locations, triage weights, and timer ratios at the application layer; low-signal zone occupancy indicators at the network layer; and battery level, wind zone proximity, and local obstacle maps at the physical layer. The action space $\mathcal{A}_t = \{\textsc{north, south, west, east, hover, land}\}$ is shared across all $N$ drones and is identical at every time step. The scalar $\alpha \in [0,1]$ trades off utilization against priority-weighted service efficiency and is treated as a design parameter whose value is reported in Section~\ref{sec:setup}. The policy $\pi^*$ is learned via a centralized Double Deep Q-Network with a QMIX-style monotonic mixing network operating under a Centralized Training with Decentralized Execution (CTDE) paradigm, as described in Section~\ref{sec:algorithm}.

\section{CEDA Solution Approach}
\label{sec:algorithm}

This section presents our deep reinforcement learning framework for cooperative medical supply delivery under dynamic and uncertain operational conditions. The method explicitly integrates environmental hazards, battery constraints, triage-driven prioritization, and inter-agent coordination into a unified, cross-layer decision-making strategy. Unlike traditional pipeline approaches that treat path planning and task assignment as separate modules, our approach embeds both concerns directly into the learning loop. This enables the policy to internalize real-time trade-offs between route efficiency, patient urgency, energy safety, and spatial coordination between agents.

\subsection{CEDA Methodology}
\label{ceda_meth}
\begin{figure*}[h]
    \centering
    \includegraphics[width=.8\linewidth]{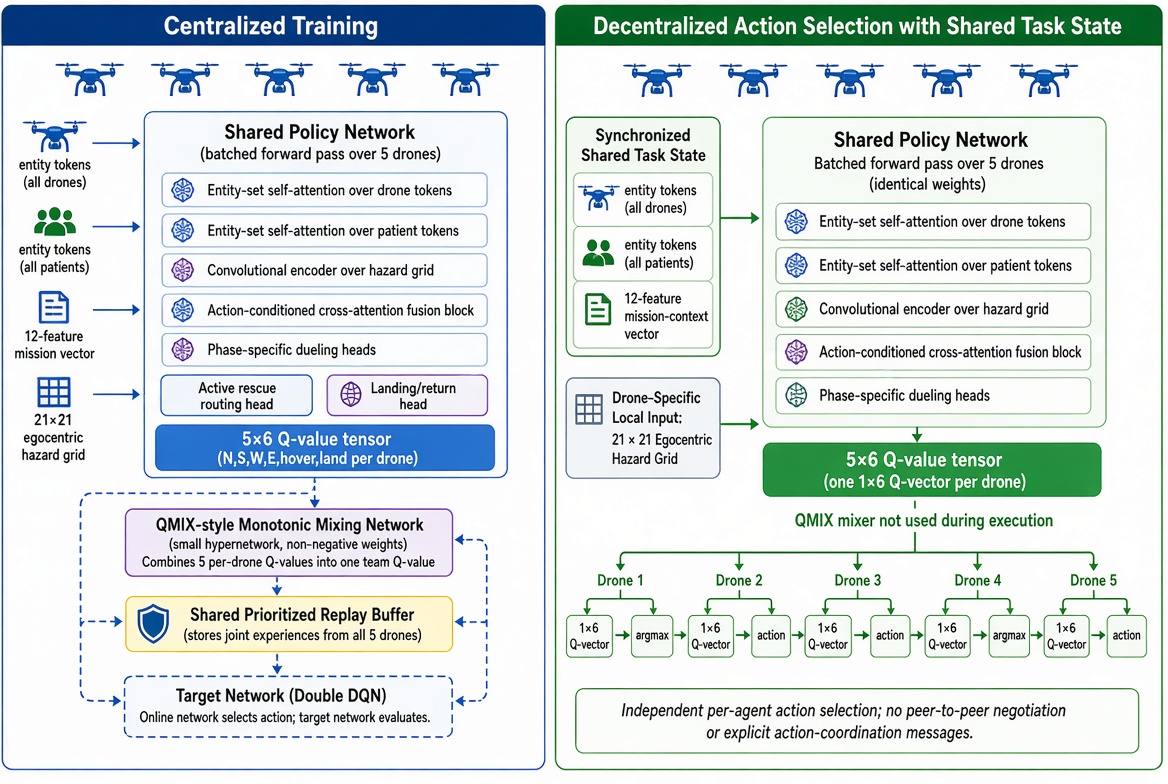}
    \caption{Centralized training and decentralized action selection with shared task-state access for cooperative medical drone delivery. During execution, the shared policy network produces one six-dimensional Q-value vector per drone, from which each drone independently selects its action.}
    \label{fig:architecture_diagram}
\end{figure*}

At the core of our approach is a Centralized Training with Decentralized Execution (CTDE) paradigm that supports cooperative multi-agent scheduling. Figure~\ref{fig:architecture_diagram} provides an overview of the architecture, illustrating how drone agents interact with the environment, share experience with a centralized trainer during training, and select their own actions independently at execution time. During training, all drone agents collectively contribute transitions to a shared replay buffer accessible to a centralized learner with full visibility of the joint state, including obstacle maps, patient locations, timer states, wind zones, low-signal zones, and the position and battery level of every agent in the fleet. This global view allows the learning algorithm to coordinate policies across agents, promote safe routing, and encode awareness of both environmental hazards and triage urgency.

Once training is complete, the shared policy network produces an individual Q-value vector for each drone, from which each drone independently selects its action, without peer-to-peer negotiation or explicit action-coordination messages at decision time. At execution, the policy uses a synchronized task-state representation comprising: (i)~entity tokens for all active drones, (ii)~entity tokens for all active patients, and (iii)~the global mission-context vector, together with the ego-drone's local hazard grid. Decentralization in CEDA therefore refers specifically to independent per-agent action selection rather than local-only observation; the QMIX-style mixing network is restricted to centralized training and is not required during execution. In the PX4 SITL implementation described in Section~\ref{sec:px4_setup}, this shared task state is updated from live vehicle telemetry through the mission-level world model before the trained network is queried.

Conventional planning methods such as dynamic programming or heuristic-based approaches often struggle in such settings, as they typically assume full state observability and static environmental models~\cite{koenig2005fast}. In contrast, our medical delivery scenario is characterized by evolving hazard zones, stochastically failing communication links, dynamically spawning patients with heterogeneous triage priorities, and actions that can silently fail to execute in low-signal zones. Reinforcement learning is well-suited to such challenges, enabling agents to learn adaptive policies from trial-and-error interaction and to make feedback-driven decisions under uncertainty.

\subsubsection{Network Architecture Overview}
\label{net_arch_over}
\begin{figure}[h]
    \centering
    \includegraphics[width=1.1\linewidth]{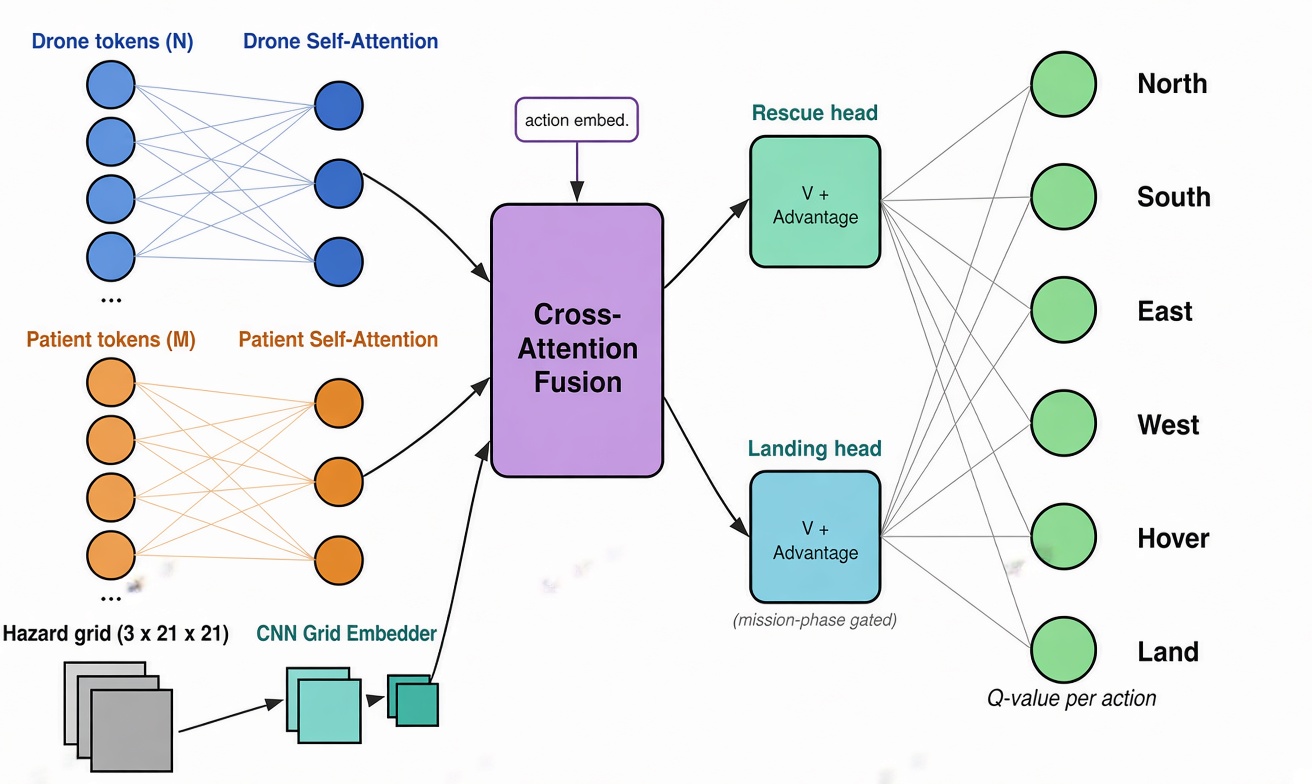}
    \caption{Neural network architecture for a single-drone forward pass through the shared policy network. Drone and patient tokens are encoded via self-attention, while the egocentric hazard grid is processed by a convolutional encoder. The three streams are fused through action-conditioned cross-attention and routed through phase-specific rescue and landing heads to produce Q-values for the six discrete actions. The same computation is applied to each drone in the fleet.}
    \label{fig:network_architecture_diagram}
\end{figure}

Figure~\ref{fig:network_architecture_diagram} illustrates the shared policy network's forward pass for a single drone. The drone and patient populations are each encoded as a set of tokens and processed by dedicated self-attention blocks, allowing the network to natively support a variable number of active drones and patients rather than a fixed-size input. In parallel, the drone's egocentric hazard grid is processed by a convolutional encoder into a compact spatial embedding. These three streams -- drone entities, patient entities, and the hazard-grid embedding -- are combined, together with an embedding of the drone's previous action, by an action-conditioned cross-attention fusion block. The fused representation feeds two phase-specific dueling value/advantage heads: a rescue head active while the drone is engaged in patient delivery, and a landing head gated on by the mission-phase indicator once the drone begins its return to base. Both heads project onto a shared six-action output layer (\textsc{north, south, west, east, hover, land}), yielding one Q-value per action for the drone.

\subsubsection{State Encoding and Feature Representation}
We model each drone $i \in \{1, \dots, N\}$ using a per-drone observation representation $o_t^i$ built from four feature categories: agent-level, patient-level, mission-level, and local environmental context.

Each drone token (22 features) encodes its position, battery, landed/died flags, landing zone position, previous move direction and collision outcome, a one-hot previous action, shortest-path distance to its landing zone, a safe-return battery margin, current wind/low-signal occupancy, and an energy-return-mode flag. Every active drone is encoded as its own token via set-attention, rather than only the ego-drone plus a single ``other agent'' feature.

Each patient token (10 features) encodes its position, remaining timer, current and initial triage weight, active/pending/delivered/died status, and an elapsed-response-time fraction, processed via a separate entity-set transformer with a null attention index for absent slots.

A 12-feature mission-context vector supplements these with global episode state: episode progress, time to next spawn, spawned/pending/delivered/died/landed fractions, an all-spawned flag, a landing-phase flag, and per-triage-class delivery target debt.

The environmental features are three local grid channels -- obstacle/boundary, wind, low-signal -- centered on the drone within a ten-cell radius, giving an egocentric hazard view for anticipating nearby hazards without global map access.

\subsubsection{Action Space}
Each drone selects from a discrete action set $\mathcal{A} = \{$\textsc{north, south, west, east, hover, land}$\}$ at every time step. The four movement actions advance the drone by one cell in the corresponding cardinal direction, subject to boundary and obstacle constraints. The \textsc{hover} action holds the drone's position in place at reduced energy cost relative to movement, and the \textsc{land} action is effective only when the drone occupies its designated landing zone; executing it elsewhere incurs a penalty without changing the drone's position. When a drone successfully lands, it is removed from active play for the remainder of the episode and its position is frozen.

In wind zones, movement actions are subject to stochastic failure: with some probability, the intended action is ignored and the drone remains stationary, consuming elevated battery at the wind-zone drain rate regardless. In low-signal zones, a similar stochastic failure applies to movement commands with probability $p_{\text{ls}} = 0.5$. These failure modes are not observable to the agent prior to action selection, requiring the policy to learn risk-aware avoidance behavior implicitly through reward experience.

\subsubsection{Reward Function Design}
\label{subsec:reward_design}
The reward function is constructed to promote timely triage-prioritized deliveries and safe navigation, while penalizing hazardous or inefficient behaviors. It decomposes into three components: a per-step reward that shapes continuous movement and routing behavior, a delivery/collision/landing reward that signals core mission-critical events, and a set of outcome-level rewards that directly score fairness and priority-ordering quality at the end of an episode.

The per-step reward for drone $i$ at time $t$ is:
\begin{multline}
\label{eq:reward_step}
r^i_{\text{step}} = -\delta 
+ \mathbb{1}[\text{clean cell}] \cdot \beta 
+ \Phi^i_t 
- \mathbb{1}[\text{wind}] \cdot \gamma_w \\
- \mathbb{1}[\text{low signal}] \cdot \gamma_s 
- \mathbb{1}[b^i_t < b_{\text{low}}] \cdot \gamma_b
- \mathbb{1}[\text{hover}] \cdot \gamma_h \\
- \frac{1}{2}\sum_{j \neq i} 
\mathbb{1}[\|pos^i_t - pos^j_t\|_1 \leq r_{\text{close}}] \cdot C
- \lambda_{\text{wait}} N_{\text{pending}}(t)
\end{multline}
\noindent where $\delta$ is a small constant time-step penalty that incentivizes efficient routing; $\beta$ rewards traversal of unobstructed, hazard-free cells; $\Phi^i_t$ is a potential-based shaping term defined as $\Phi^i_t = \Lambda \cdot (D^i_{t-1} - D^i_t)$, where $D^i_t$ is the Manhattan distance from drone $i$ to its nearest undelivered patient (or to its landing zone if all patients are delivered) and $\Lambda$ is a shaping scale factor; $\gamma_w$ and $\gamma_s$ penalize entry into wind and low-signal zones respectively; $\gamma_b$ penalizes operation below the low-battery threshold $b_{\text{low}}$; $\gamma_h$ penalizes the \textsc{hover} action to discourage idling; $C$ penalizes spatial proximity between any pair of active drones within radius $r_{\text{close}}$, summed over all such pairs in the fleet; and $\lambda_{\text{wait}} N_{\text{pending}}(t)$ applies a class-independent response-wait penalty proportional to the number of patients that remain pending at time $t$, encouraging timely service without changing their relative triage priorities. Two additional routing-awareness penalties discourage drones from selecting undelivered patients whose shortest path is dominated by wind or low-signal cells, and an energy-aware navigation term rewards battery-conscious route choices; exact parameter values are reported in Table~\ref{tab:reward_parameters}.


The delivery, collision, and landing reward captures discrete events:

\begin{equation}
\label{eq:reward_milestone}
\begin{split}
r^i_{\text{event}} &= \mathbb{1}[\text{delivery}_p] \cdot R_{\text{goal}} 
\cdot \frac{t_p^{\text{rem}}}{T_{\max}} \cdot w_p(t) + \mathbb{1}[\text{delivery}_p] \cdot R_{\text{response}} \\
&\quad + \mathbb{1}[\text{delivery}_p] \cdot \Lambda_{\text{svc}} \cdot \max\!\left(0,\ \Psi_t - \Psi_{t-1}\right) \\
&\quad - \mathbb{1}[\text{collision}] \cdot R_{\text{crash}} \\
&\quad - \mathbb{1}[\text{battery depleted}] \cdot R_{\text{bat}} \\
&\quad + \mathbb{1}[\text{landed}] \cdot R_{\text{land}} \\
&\quad - \mathbb{1}[\text{death}_p] \cdot P_{\text{death}} \cdot \left(1 + \kappa(w_p^{0} - 1)\right)
\end{split}
\end{equation}
\noindent where $R_{\text{goal}}$ is the base delivery reward scaled by the timer ratio and the patient's current triage weight $w_p(t)$ (Section~\ref{sec:triage}); $R_{\text{response}}$ is a fixed bonus for successful delivery; $\Psi_t = \sum_{\rho \in \{1,2,3\}} \frac{\rho}{\rho_{\max}} \cdot \frac{\min(D_\rho(t), G_\rho)}{M}$ is a priority-weighted service-completion potential, where $D_\rho(t)$ is the number of class-$\rho$ patients delivered by time $t$, $G_\rho$ is the per-class delivery target, $\rho_{\max}$ is the maximum triage weight, and $M$ is the maximum patient count; the reward is proportional to the non-negative increase in $\Psi_t$ at each delivery, scaled by $\Lambda_{\text{svc}}$, so only forward progress in weighted service completion is rewarded, distinct from the raw-distance potential $\Phi^i_t$ in Eq.~(\ref{eq:reward_step}); $R_{\text{crash}}$ penalizes obstacle and inter-agent collisions; $R_{\text{bat}}$ penalizes battery depletion; $R_{\text{land}}$ rewards successful landing at the designated zone, with a separate penalty for landing attempts elsewhere; and $P_{\text{death}}$ is a shared penalty incurred when patient $p$'s timer expires without delivery, scaled by $\kappa$ and the patient's initial triage weight $w_p^{0}$ so that expiration of a patient with higher initial priority incurs a proportionally larger penalty.


In addition to these per-step and per-event terms, we introduce a set of outcome-level rewards that directly score fairness and priority-ordering quality rather than relying solely on indirect shaping. At each step, a fairness outcome reward $R_{\text{fair}} \cdot f_t$ is computed from a running fairness score $f_t$ over per-class service rates, a triage-ordering outcome reward $R_{\text{order}} \cdot o_t$ is computed from the fraction of pairwise Critical-before-Stable service orderings satisfied, and a triage response-time outcome reward further rewards maintaining a favorable Critical-versus-Stable response-time gap.

The total reward at each step is:
\begin{equation}
r^i_t = \text{clip}(r^i_{\text{step}},\ -\delta_{\max},\ \delta_{\max}) + r^i_{\text{event}} + r^i_{\text{outcome}}
\end{equation}
The step reward is clipped to prevent large shaping gradients from destabilizing training, while event and outcome rewards are applied unclipped to preserve their full behavioral signal.

\subsection{Centralized Q-Network Architecture}
\label{sec:dqn_architecture}
Given the discrete action structure, we adopt a value-based formulation for CEDA: Double DQN provides direct per-action Q-value evaluation, parameter sharing across homogeneous drones, and compatibility with the QMIX-style mixing network used for cooperative credit assignment during centralized training. Section~\ref{subsec:baselines} evaluates MAPPO as a representative policy-based MARL baseline, providing a direct comparison against this value-based formulation.

At the core of our framework is a shared policy network $Q_\theta$ that computes per-drone action-value vectors from the entity-attention encoding described in Section~\ref{net_arch_over}. Rather than being queried once per agent, $Q_\theta$ processes the entire fleet in a single batched forward pass, producing a $(N, |\mathcal{A}|)$ tensor of Q-values, one $|\mathcal{A}|=6$-dimensional vector per drone, fused from two phase-specific dueling value/advantage heads, one specialized for active-rescue routing and one for end-of-mission return/landing behavior. Because the underlying entity encoding is permutation-invariant over drone and patient tokens, the same network parameters apply regardless of fleet size $N$ or active patient count.

A separate target network $Q_{\theta^-}$ with identical architecture is maintained from $Q_\theta$ every $K$ learner (gradient) steps. To reduce Q-value overestimation, action selection and evaluation are decoupled following Double DQN: for each transition, the \emph{online} network $Q_\theta$ selects each drone's greedy next action (with invalid actions, such as an out-of-bounds move or a \textsc{land} attempted outside the designated zone, masked out before the arg max), and the \emph{target} network $Q_{\theta^-}$ evaluates the value of that selected action.

Because the fleet's individual action-values must be combined into a single team objective under CTDE, a QMIX-style mixing network $Q_{\text{tot}} = f_{\text{mix}}(Q_\theta^0, \ldots, Q_\theta^{N-1};\ s_t)$ combines the $N$ masked, per-drone chosen Q-values into a joint team value. The mixer is a small hypernetwork conditioned on a centralized state encoding $s_t$ (built from the same joint observation available during centralized training), producing strictly non-negative mixing weights via a bounded sigmoid gate, followed by a state-dependent bias term. Constraining the mixing weights to be non-negative guarantees that $Q_{\text{tot}}$ is monotonically non-decreasing in each per-drone value, which is precisely the Individual-Global-Max (IGM) property invoked in Section~\ref{ceda_meth} to justify independent, negotiation-free action selection at execution time. A separate target mixer, hard-copied alongside $Q_{\theta^-}$, is used identically to $Q_{\theta^-}$ in the target computation below.

Experience replay is implemented via a prioritized joint replay buffer (capacity $|\mathcal{B}|$, prioritization exponent $\alpha$, and importance-sampling exponent $\beta$ annealed over training, alongside $n$-step returns) that stores transitions of the form $(o_t^0,\ldots,o_t^{N-1}, a_t^0,\ldots,a_t^{N-1}, r_t^0,\ldots,r_t^{N-1}, o_{t+1}^0,\ldots,o_{t+1}^{N-1}, \text{done})$. Mini-batches of size $B$ are sampled proportionally to transition priority from the buffer, after an initial replay warmup period, to compute the training loss.

Two complementary temporal-difference losses are combined at each update. A \emph{team} loss trains the mixer and policy network jointly against an $n$-step, $\gamma$-discounted team-level reward target using $Q_{\text{tot}}$ from the target network and target mixer; and a \emph{local} loss trains each drone's own chosen utility directly against a locally-attributed reward target (isolating, e.g., each drone's own potential-shaping and routing contributions), masked to exclude drones that have already landed or died and averaged over active drones. Both losses use the smooth $L_1$ (Huber) loss, and the team-level temporal-difference error is used to update transition priorities in the replay buffer. Let $\delta^i_{\text{team}}$ and $\delta^i_{\text{local}}$ denote the team- and local-level TD errors for drone $i$; the combined loss is:

\begin{multline}
\label{eq:loss}
\mathcal{L}(\theta) = \mathbb{E}_{(s,a,r,s') \sim \mathcal{B}}
\left[ w \cdot \ell_{\text{Huber}}\!\left(\delta_{\text{team}}\right) \right] \\
+ \lambda_{\text{local}} \cdot \mathbb{E}_{(s,a,r,s') \sim \mathcal{B}}
\left[ w \cdot \frac{1}{|\mathcal{A}_{\text{active}}|}\sum_{i \in \mathcal{A}_{\text{active}}} \ell_{\text{Huber}}\!\left(\delta^i_{\text{local}}\right) \right]
\end{multline}
\noindent where $w$ is the per-sample prioritized-replay importance-sampling weight, $\mathcal{A}_{\text{active}}$ is the set of drones neither landed nor deceased in that transition, and $\lambda_{\text{local}}$ weights the local loss relative to the team loss. Gradients are clipped to a fixed norm before each parameter update to prevent instability during early training. The Adam optimizer is used with a fixed learning rate and discount factor $\gamma$. The training procedure is described formally in Algorithm~\ref{alg:ctde_dqn}.

\begin{algorithm}[t]
\caption{CEDA for Cooperative Medical Drone Delivery}
\label{alg:ctde_dqn}
\small
\begin{algorithmic}[1]
\STATE \textbf{Input:} Learning rate, discount factor $\gamma$, exploration rate $\epsilon$, target update frequency $K$
\STATE Initialize policy network $Q_\theta$, mixer $f_{\text{mix}}$, and target copies $Q_{\theta^-}, f_{\text{mix}}^-$
\STATE Initialize prioritized joint replay buffer $\mathcal{B} \leftarrow \emptyset$
\FOR{episode $= 1$ \textbf{to} $E$}
    \STATE Reset environment with $N$ drones; sample patient survival profiles $(a_i, b_i, \theta_i^{\text{serious}}, \theta_i^{\text{critical}})$ per Eq.~(\ref{eq:survival}--\ref{eq:triage_escalation})
    \STATE Observe initial joint state $[o_1^0,\ldots,o_1^{N-1}]$ across physical, network, and application layers

    \FOR{$t = 1$ \textbf{to} $T_{ep}$}
        \STATE $\{a_t^i\}_{i=1}^{N} = \epsilon\text{-greedy}(Q_\theta(o_t^0,\ldots,o_t^{N-1}))$, masking invalid actions
        \STATE Execute all $a_t^i$; apply stochastic failure in wind/low-signal zones; update $w_i(t)$ per Eq.~(\ref{eq:triage_escalation})
        \STATE Compute $\{r_t^i\}$ per Eqs.~(\ref{eq:reward_step})--(\ref{eq:reward_milestone}) and outcome rewards (Section~\ref{subsec:reward_design})
        \STATE Set \textsc{terminal} $\leftarrow$ all drones landed/depleted \textbf{or} episode horizon reached
        \STATE Store transition $(o_t^{0:N-1}, a_t^{0:N-1}, r_t^{0:N-1}, o_{t+1}^{0:N-1}, \textsc{terminal})$ in $\mathcal{B}$ with maximal priority
        \IF{$|\mathcal{B}| \geq $ warmup size}
            \STATE Sample prioritized mini-batch; compute $\mathcal{L}(\theta)$ per Eq.~(\ref{eq:loss}) (Double DQN target via $Q_\theta, Q_{\theta^-}, f_{\text{mix}}^-$); update $\theta$ via Adam; update replay priorities
        \ENDIF
        \IF{\textsc{terminal}} \STATE \textbf{break} \ENDIF
        \IF{learner step $\bmod\ K = 0$} \STATE $Q_{\theta^-} \leftarrow Q_\theta$;\ $f_{\text{mix}}^- \leftarrow f_{\text{mix}}$ \ENDIF
    \ENDFOR
    \STATE $\epsilon \leftarrow \max(\epsilon_{\min},\ \epsilon \cdot \epsilon_{\text{decay}})$
\ENDFOR
\normalsize
\end{algorithmic}
\end{algorithm}

At the start of each episode, the environment is reset with randomized patient positions, weights, and hazard zone layouts to promote generalization. Actions are selected via an $\epsilon$-greedy policy, with $\epsilon$ annealed from $\epsilon_{\text{start}}$ to $\epsilon_{\text{end}}$ over the first 95\% of training episodes. At execution, the mixer and target network are not used, and each drone selects its action independently as described in Section~\ref{ceda_meth}.

\subsection{Complexity Analysis}
\label{complexity}
The computational cost of the shared policy network is governed primarily by the entity-attention encoders. For the patient and drone token sets, the self-attention operations scale as $\mathcal{O}(M^2 d_e)$ and $\mathcal{O}(N^2 d_e)$, respectively, where $M$ and $N$ denote the numbers of active patients and drones and $d_e$ is the entity-embedding dimension. Accordingly, the architecture can process variable numbers of active entities without changing the policy-network dimensionality, while the computational cost increases with the size of the entity sets. Further, parameter sharing avoids maintaining a separate policy network for each drone and keeps the number of learned policy parameters independent of fleet size. The QMIX-style mixing network introduces additional training-time computation but is not used during execution.

In addition to the computational cost of the attention layers, memory requirements are influenced primarily by the prioritized joint replay buffer, which stores full joint transitions and maintains a sum-tree structure for prioritized sampling. The resulting storage requirement increases with the replay-buffer capacity, fleet size, and per-agent observation size, while priority updates and sampling require $\mathcal{O}(\log|\mathcal{B}|)$ operations. Consequently, parameter sharing reduces model-storage and synchronization requirements relative to independent learner architectures, although the replay-buffer and attention costs continue to increase with the number of active entities. At deployment, the shared policy network produces one Q-value vector for each drone from the synchronized task-state representation, and each drone independently selects its action from the corresponding vector.

\section{Performance Evaluation}
\label{sec:results}

In this section, we evaluate the proposed CEDA for cooperative medical drone delivery under dynamic and uncertain operational conditions. Our goal is to demonstrate how the jointly trained policy learns to navigate hazardous environments, coordinate between agents, prioritize triage-critical patients, and preserve service fairness across all patient weight classes. We present training convergence analysis, safety and hazard avoidance results, operational efficiency metrics, and a detailed triage fairness evaluation grounded in the Priority-Preserving Fair Scheduling mechanisms described in Section~\ref{sec:fairness}.

\subsection{Simulation Setup and Experimental Configuration}
\label{sec:setup}
We consider a disaster response scenario involving 5 autonomous medical delivery drones operating over a discrete grid environment with static obstacles, dynamic wind zones, and dynamic low-signal zones. All agents are trained using the CEDA framework described in Section~\ref{sec:algorithm}.

\textit{Environment Settings:} The simulation grid spans $100 \times 100$ cells populated with 800 static obstacles. Dynamic wind zones and low-signal zones are refreshed every 500 time steps along A*-computed inter-patient paths to maximally stress the agents' routing decisions. Five drone agents are initialized from fixed start positions with full battery and must return to dedicated landing zones upon mission completion.

\textit{Patient Settings:} Up to $M = 50$ patients spawn dynamically throughout each episode, beginning with $N_{\text{init}} = 20$ active patients at episode start. New patients spawn in batches of 4 to 6 at intervals of $60 \pm 5$ steps, with spawning ending at step 400. Each patient is assigned a triage weight $w_i$ drawn uniformly from $\mathcal{W} = \{1, 2, 3\}$ corresponding to Stable, Urgent, and Critical urgency levels, and carries a countdown timer initialized to 220 steps in the final curriculum stage used to produce the reported results; the constant $T_{\max} = 300$ is retained as the fixed normalization denominator in the timer ratio $t_i^{\text{rem}}/T_{\max}$ appearing in the state representation and reward function (Equation~4), rather than as the patient's actual deadline. Per-patient survival curve parameters are sampled conditioned on initial triage level: Stable patients receive $a_i \in [0.02, 0.05]$ and $b_i \in [3.0, 5.0]$; Urgent patients receive $a_i \in [0.05, 0.10]$ and $b_i \in [2.0, 3.5]$; Critical patients receive $a_i \in [0.10, 0.20]$ and $b_i \in [1.0, 2.5]$. Escalation thresholds are sampled as $\theta_i^{\text{serious}} \in [0.40, 0.70]$ and $\theta_i^{\text{critical}} \in [0.10, 0.30]$, subject to $\theta_i^{\text{critical}} < \theta_i^{\text{serious}} - 0.05$.

Our fleet size and patient counts are grounded in prior disaster-response studies based on EMS1 survey data~\cite{guerdan2017augmented,jiang2019hierarchical}. These studies report 1{,}746 survey responses, primarily from first responders, and indicate that 60\% of disaster incidents involve between 5 and 30 triage patients. Our evaluation of $M=50$ patients exceeds the upper bound reported by the large majority of survey respondents.

\textit{Network Settings:} The shared policy network follows the entity-attention architecture, encoding a variable number of drone and patient entities and producing Q-values over the six-action space. Centralized training additionally employs a QMIX-style mixing network, described in Section~\ref{sec:dqn_architecture}, to combine per-drone Q-values into a joint team value. The model is trained using the PyTorch framework with the Adam optimizer, following the training procedure formalized in Algorithm~\ref{alg:ctde_dqn}. All training hyperparameters are summarized in Table~\ref{tab:hyperparams}.

\begin{table}[h]
\centering
\caption{Training hyperparameters for the CEDA framework}
\label{tab:hyperparams}
\small
\begin{tabular}{@{}lc@{}}
\toprule
\textbf{Hyperparameter} & \textbf{Value} \\
\midrule
Training episodes          & 12,000 \\
Max steps per episode      & 800 \\
Epsilon schedule $\epsilon_{\text{start}} \to \epsilon_{\text{mid}} \to \epsilon_{\text{end}}$ & $1.0 \to 0.15 \to 0.01$ \\
Curriculum epsilon reset / floor & $0.25 \to 0.03$ over 750 ep. \\
Replay buffer capacity     & 500,000 \\
Replay warmup              & 50,000 transitions \\
Prioritization exponent $\alpha_{\text{PER}}$ & 0.4 \\
Importance-sampling exponent $\beta$ & $0.4 \to 1.0$ \\
$n$-step return length     & 5 \\
Batch size                 & 256 \\
Discount factor $\gamma$   & 0.995 \\
Utilization--triage trade-off $\alpha$ & $0.5$ \\
Learning rate              & $5 \times 10^{-5}$ \\
Gradient clip norm         & 7.5 \\
Local TD loss weight $\lambda_{\text{local}}$ & 0.25 \\
Target update frequency    & 625 learner steps \\
Action space $|\mathcal{A}|$ & 6 \\
Fleet size $N$             & 5 \\
Max patients $M$           & 50 \\
Initial patients $N_{\text{init}}$ & 20 \\
Patient timer norm. constant $T_{\max}$ & 300 steps \\
Patient countdown timer (final training) & 220 steps \\
Patient spawn batch size   & 4--6 patients \\
Patient spawn interval     & $60 \pm 5$ steps, until step 400 \\
Triage weights $\mathcal{W}$ & $\{1, 2, 3\}$ \\
Closeness radius $r_{\text{close}}$ & 4 cells \\
Grid size                  & $100 \times 100$ \\
Static obstacles           & 800 \\
Hazard zone refresh interval & 500 steps \\
\bottomrule
\normalsize
\end{tabular}
\end{table}

\begin{table}
\centering
\caption{Summary of reward parameters used by the CEDA framework}
\label{tab:reward_parameters}
\footnotesize
\setlength{\tabcolsep}{3pt}
\renewcommand{\arraystretch}{1.08}

\begin{tabularx}{\columnwidth}{@{}
    >{\raggedright\arraybackslash}p{0.25\columnwidth}
    >{\raggedright\arraybackslash}X@{}}
\toprule
\textbf{Component} & \textbf{Parameters and values} \\
\midrule

Step shaping
&
Time-step penalty $\delta=0.02$;
clean-cell reward $\beta=0.01$;
hover penalty $\gamma_h=1.0$.
\\
\addlinespace[2pt]

Patient shaping
&
Navigation-distance potential scale $\Lambda=0.5$;
priority-weighted service-completion scale $\Lambda_{\text{svc}}=300.0$.
\\
\addlinespace[2pt]

Wind avoidance
&
Occupancy penalty $=3.0$;
entry penalty $=4.0$;
dominated-route penalty $=6.0$;
shortcut-route penalty $=3.0$.
\\
\addlinespace[2pt]

Low-signal avoidance
&
Occupancy penalty magnitude $\gamma_s=4.0$;
entry penalty magnitude $=5.0$;
dominated-route penalty magnitude $=8.0$;
shortcut-route penalty magnitude $=4.0$.
\\
\addlinespace[2pt]

Energy management
&
Low-battery penalty $\gamma_b=0.1$ below
$b_{\text{low}}=20.0$;
energy-usage penalty $=0.05$;
safe-return reserve penalty $=3.0$;
energy-margin shaping factor $=5.0$;
battery-depletion penalty $R_{\text{bat}}=100.0$.
\\
\addlinespace[2pt]

Coordination and safety
&
Agent closeness penalty $C=0.5$ within
$r_{\text{close}}=4$ cells;
obstacle/agent collision penalty
$R_{\text{crash}}=45.0$.
\\
\addlinespace[2pt]

Delivery and response
&
Base delivery reward $R_{\text{goal}}=10.0$;
fixed successful-delivery bonus $R_{\text{response}}=12.0$;
pending-patient response-wait penalty $\lambda_{\text{wait}}  =0.015$ per patient.
\\
\addlinespace[2pt]

Landing and patient loss
&
Successful landing reward $R_{\text{land}}=30.0$;
invalid landing attempt penalty $=300.0$;
patient death penalty $P_{\text{death}}=30.0$;
priority growth $\kappa=0.25$.
\\
\addlinespace[2pt]

Fairness and priority
&
Fairness outcome reward $=50.0$;
triage-ordering outcome reward $=50.0$;
triage-response outcome reward $=100.0$.
\\

\bottomrule
\end{tabularx}
\normalsize
\end{table}

\subsection{Reward Function Parameters}
\label{subsec:reward_params}
Table~\ref{tab:reward_parameters} summarizes the complete set of reward parameters used during training, grouped by their functional roles. In particular, the hazard-related terms include entry, occupancy, and route-selection penalties for wind and low-signal regions. All quantities identified as penalties are reported as positive magnitudes in Table~\ref{tab:reward_parameters}. Accordingly, the corresponding negative signs are applied explicitly in the reward formulation. Further, the priority-weighted service-completion potential $\Psi_t$ (Eq.~\ref{eq:reward_milestone}) rewards progress in weighted delivery completion at each delivery event, distinct from the raw-distance navigation potential $\Phi^i_t$ used in the per-step shaping term, while the patient-death penalty scales with the patient's initial triage weight.

\subsection{Reward Convergence and Mission Success}
\label{subsec:convergence}

\begin{figure*}[h]
    \centering
    \includegraphics[width=0.95\linewidth]{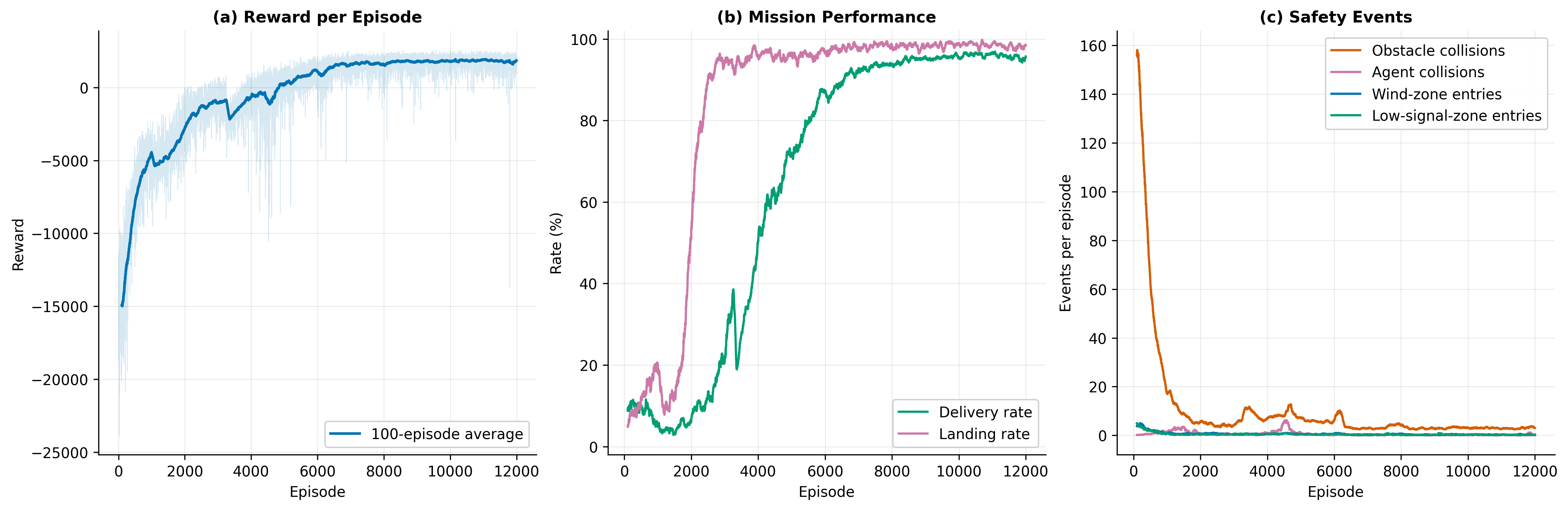}
    \caption{Training and operational performance over 12,000 episodes for the five-drone, fifty-patient configuration. (a) Total reward per episode with a 100-episode moving average. (b) Delivery and landing rates. (c) Obstacle collisions, inter-agent collisions, wind-zone entries, and low-signal-zone entries per episode.}
    \label{fig:reward_learning}
\end{figure*}

Figure~\ref{fig:reward_learning} presents the overall learning trajectory across 12,000 episodes for the expanded five-drone, fifty-patient configuration. The total reward (a) rises from approximately $-15{,}000$ to a positive plateau near $1{,}800$ in late training, indicating convergence of the average reward. Mission performance (b) shows the landing rate reaching approximately 95\% around episode 4,000 and settling near 98\% in late training, while the delivery rate reaches approximately 95--96\% by episode 12,000. Safety events (c) show obstacle collisions falling sharply to single digits by episode 2,000, with brief re-elevations during training, while inter-agent collisions remain low. Wind-zone entries and low-signal-zone entries decline to well below one entry per episode on average in late training, indicating reduced exposure to both types of hazard.

\subsection{Triage Fairness and Priority-Preserving Scheduling}
\label{subsec:triage}
\begin{figure*}[h]
    \centering
    \includegraphics[width=0.95\linewidth]{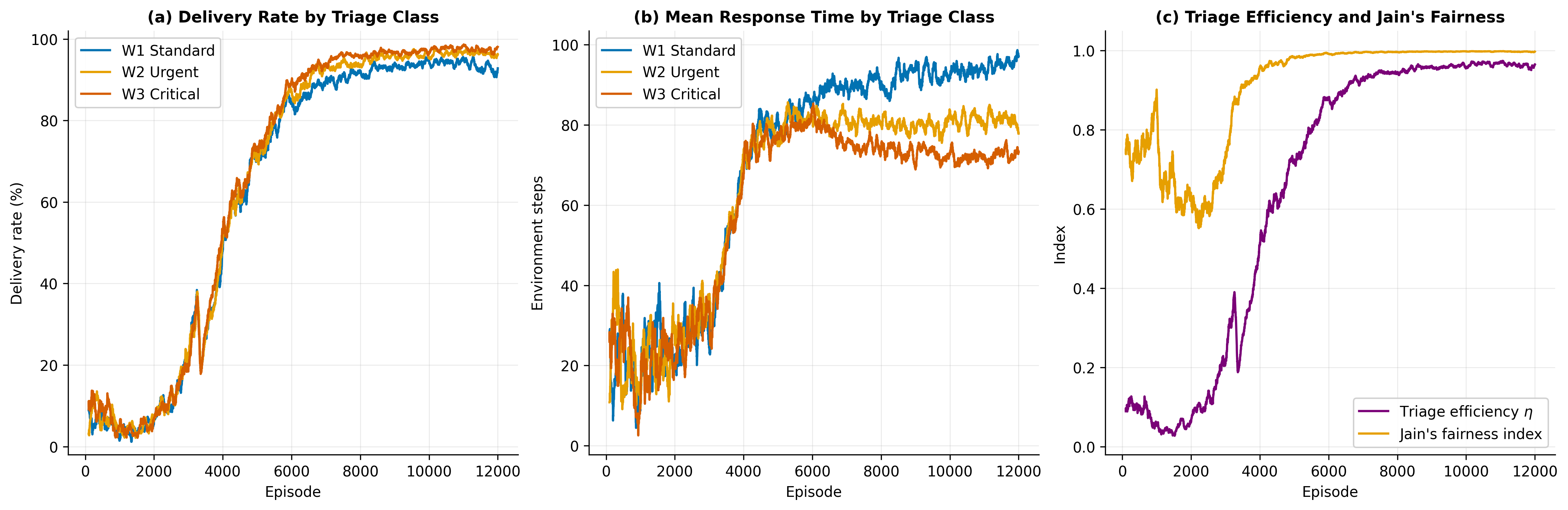}
    \caption{Triage performance and fairness metrics over 12,000 training episodes. (a) Delivery rate by initial triage class. (b) Mean response time from patient spawn to successful delivery by initial triage class, measured in environment steps. (c) Triage efficiency $\eta$ and Jain's Fairness Index across triage-class delivery rates.}
    \label{fig:triage}
\end{figure*}

Figure~\ref{fig:triage} evaluates whether the Priority-Preserving Fair Scheduling mechanisms produce the intended two-part fairness criterion: that clinical priority ordering is preserved, and that no lower-priority class is abandoned through starvation. CEDA satisfies both conditions at the expanded five-drone, fifty-patient scale.

Figure~\ref{fig:triage} (a) shows that delivery rates reach approximately 92--98\% for all three triage classes by episode 12,000, with slightly higher rates for Urgent and Critical patients than for Stable patients. Further, Figure~\ref{fig:triage}(b) shows a clear priority ordering in response time. The smoothed mean response times in late training are approximately 73, 78, and 97 environment steps for Critical, Urgent, and Stable patients, respectively. To examine variability beyond the training curves, we evaluated per-patient response times over 100 held-out episodes comprising 4{,}801 successful deliveries at the greedy policy setting ($\epsilon=0$). The mean $\pm$ SD response times are $70.3 \pm 51.2$ steps for Critical patients ($n=1{,}633$), $78.7 \pm 56.2$ steps for Urgent patients ($n=1{,}605$), and $92.6 \pm 58.7$ steps for Stable patients ($n=1{,}563$). The corresponding medians are 56.0, 66.0, and 86.0 steps. Thus, both the mean and median response times preserve the Critical~$<$~Urgent~$<$~Stable ordering, while the reported standard deviations capture the within-class variation in service delay. Finally, Figure~\ref{fig:triage} (c) shows triage efficiency $\eta$ stabilizes near 0.95--0.97, while Jain's Fairness Index converges to approximately 1.0. Together, these results support priority-sensitive service while maintaining comparable service levels across triage classes.

\subsection{Computational Complexity}
\label{subsec:complexity}

For the five-drone, fifty-patient configuration, per-agent inference averages 1.37~ms, with environment-step computation adding 2.15~ms for a total decision latency of 3.53~ms per step, indicating low computational latency on the reference hardware. Full training completes in approximately 12.5~hours over 12,000 episodes (3.76~s per episode on average), establishing computational feasibility at this scale. The shared entity-attention architecture accommodates a variable number of drones and patients without requiring a separate policy network per fleet size, and parameter sharing keeps the learned policy size independent of fleet size; the attention computation itself, however, still grows with the number of active entities, as discussed in Section~\ref{complexity}. Table~\ref{tab:marl-baselines} additionally reports architecture-level computational cost relative to contemporary MARL baselines; further empirical results, including training-time, memory, and workload figures, are provided in Supplementary Section~S1.

\subsection{Comparison Against Baselines}
\label{subsec:baselines}
To contextualize CEDA's performance, we benchmark it against four contemporary MARL baselines spanning value-decomposition and centralised-critic actor--critic approaches, trained using the same environment, reward function, curriculum schedule, training budget, and evaluation protocol, while retaining their method-specific learning architectures. After training, all five policies are frozen and evaluated over the same $N_{\mathrm{eval}}=100$ previously unseen environment seeds (governing patient spawn locations, triage assignments, and hazard-zone layouts) using deterministic/greedy action selection, ensuring a fair head-to-head comparison on identical task instances.

\subsubsection{Baselines}
\begin{itemize}
  \item \textbf{VDN}~\cite{sunehag2018vdn} --- value decomposition with an additive factorisation $Q_{\text{tot}}(\bm{\tau},\bm{a}) = \sum_{i} Q_i(\tau_i, a_i)$ and double-DQN targets. This is also the direct architectural ablation of CEDA's monotonic mixing network, with the mixer replaced by a sum.
  \item \textbf{QMIX}~\cite{rashid2018qmix} --- a monotonic mixing network with flat-state hypernetworks ($|w|=\mathrm{abs}$), i.e.\ CEDA without the entity-attention state and utility encoders. The gap between QMIX and CEDA therefore measures what those encoders contribute over vanilla monotonic mixing.
  \item \textbf{MADDPG}~\cite{lowe2017maddpg} --- decentralised actors with per-agent centralised critics $Q_i(\bm{x}, a_1,\dots,a_N)$; the discrete drone action space is handled with the straight-through Gumbel--Softmax estimator, following the discrete-action experiments in the original paper. Soft target updates; the critic is updated once per two environment steps to amortise the $N$ critic updates per learning step.
  \item \textbf{MAPPO}~\cite{yu2022mappo} --- on-policy actor--critic with a centralised value function, using GAE and the standard clipped surrogate objective.
\end{itemize}
All methods share CEDA's discount factor, reward scaling, and $\epsilon$-schedule (where applicable).

\begin{table*}[t]
\centering
\caption{CEDA versus contemporary MARL baselines on the full mission. Task-performance metrics are means over $N_{\mathrm{eval}}=100$ held-out evaluation episodes using identical environment seeds, frozen networks, and deterministic actions across all methods. Computational metrics are measured on the same hardware (1$\times$ NVIDIA L40S GPU, AWS EC2, Linux x86-64, CUDA). $\uparrow/\downarrow$ indicates that higher/lower is better.}
\label{tab:marl-baselines}
\begin{tabular}{lccccc}
\toprule
Metric & VDN & QMIX & MADDPG & MAPPO & \textbf{CEDA (Proposed)} \\
\midrule
\multicolumn{6}{l}{\textit{Task performance}} \\
Delivery rate $\uparrow$            & 0.65 & 0.54 & 0.62 & 0.94 & \textbf{0.96} \\
Triage efficiency $\eta$ $\uparrow$ & 0.50 & 0.57 & 0.46 & 0.77 & \textbf{0.95} \\
Mission success $\uparrow$          & 0.71 & 0.68 & 0.61 & 0.90 & \textbf{0.92} \\
Patient deaths $\downarrow$         & 4.56 & 3.95 & 5.21 & 2.79 & \textbf{2.18} \\
Drones landed $\uparrow$            & 3.94 & 3.31 & 3.23 & 4.39 & \textbf{4.75} \\
Obstacle collisions $\downarrow$    & 21.9 & 17.6 & 19.3 & 11.9 & \textbf{9.9} \\
Agent collisions $\downarrow$       & 0.57 & 0.38 & 1.02 & 0.3 & \textbf{0.3} \\
Episode length                      & 494  & 497  & 476  & 470  & \textbf{385} \\
\midrule
\multicolumn{6}{l}{\textit{Computational cost}} \\
Trainable parameters (M) $\downarrow$
  & \textbf{1.38} & 1.52 & 9.38 & 2.41 & 1.77 \\
Wall-clock training time (h) $\downarrow$
  & \textbf{10.4} & 10.7 & 11 & 14.2 & 12.5 \\
Per-agent inference (ms) $\downarrow$
  & \textbf{0.07}$^\ddagger$ & \textbf{0.07}$^\ddagger$ & 0.28 & 1.62 & 1.37 \\
Single training update (ms) $\downarrow$
  & \textbf{8.9} & 13.3 & 34.4 & 92.6 & 79.8 \\
Policy parameter memory (MB) $\downarrow$
  & \textbf{5.53} & \textbf{5.53} & 22.35 & 9.14 & 6.77 \\
Peak process memory (MB) $\downarrow$
  & \textbf{4\,688} & 5\,370 & 5\,129 & 13\,850 & 11\,397 \\
\bottomrule
\multicolumn{6}{p{0.95\linewidth}}{\footnotesize
$^\ddagger$VDN/QMIX evaluate a single shared Q-network in one batched forward for all five agents
(0.33\,ms total); the per-agent figure is that cost amortised.}
\end{tabular}
\end{table*}

\subsubsection{Results}
Table~\ref{tab:marl-baselines} shows that CEDA leads every task-performance metric, matching only MAPPO's agent-collision rate. MAPPO is the strongest baseline (triage efficiency 0.77, delivery rate 0.94, mission success 0.90), suggesting on-policy learning with a centralised value function is a competitive approach here, but it still trails CEDA on every metric while requiring more steps per episode (470 vs.\ 385), so CEDA both performs better and completes missions faster.

VDN and QMIX diverge informatively: VDN achieves higher delivery rate, mission success, and landing count, while QMIX's monotonic mixer yields fewer deaths, obstacle collisions, and agent collisions, suggesting monotonic credit assignment improves safety without improving raw throughput, and that CEDA's advantage over both is not attributable to the mixing mechanism alone. MADDPG is the weakest performer overall, which may reflect the difficulty its per-agent centralised critics face in regressing a stable joint action-value estimate against the other agents' non-stationary policies during training. No baseline recovers CEDA's landing discipline (3.2--4.4 of 5 drones landed, versus 4.75) or its delivery and triage performance, suggesting CEDA's advantage is attributable at least in part to its learned policy and architecture, entity-attention encoders and monotonic credit assignment, rather than to the environment, reward, or training protocol, all held fixed across methods.

VDN and QMIX are the cheapest baselines on every per-step metric (lowest inference time, fastest training update, smallest policy memory) and fastest to train overall (10.4~h, 10.7~h), consistent with lacking entity-attention encoders. MADDPG has by far the largest parameter count (9.38~M, 4--7$\times$ every other method) from its per-agent centralised critics, though its training time (11.0~h) stays close to the value-decomposition methods. MAPPO is the most computationally expensive baseline (highest inference time, slowest training update, largest peak memory, longest training time at 14.2~h), reflecting a genuine compute-performance trade-off given its status as the strongest baseline. CEDA's per-step costs (1.37~ms inference, 79.8~ms training update) and training time (12.5~h) fall between the value-decomposition methods and MAPPO, consistent with its entity-attention architecture being costlier per step than a flat network but cheaper than MAPPO's centralised on-policy setup.

\subsection{Cross-Layer Stress Interaction}
\label{subsec:crosslayer}

We additionally evaluated CEDA under a $3\times3$ factorial combination of network-layer stress (low-signal coverage and command failure probability) and application-layer stress (patient load and deadline tightness), with each cell evaluated over 100 episodes. The results show that application-layer load has the dominant effect on triage efficiency, reducing $\eta$ by 26--30\% from light to heavy stress, while network stress alone changes $\eta$ by at most 0.05. Network stress instead primarily affects mission completion: landing rate falls from 1.00 to 0.86 under baseline application load as network stress increases, and further to 0.69 under heavy application load, driven by battery cost from hazard-avoidance detours rather than command failures directly. The complete factorial results and heatmaps are provided in Supplementary Section~S2.

\subsection{Cross-Layer Information Awareness Ablation}
\label{subsec:awareness}

To isolate the contribution of each information layer to overall system performance, we conduct a cross-layer awareness ablation in which each layer's features are selectively removed from the agent's state vector at evaluation time while the policy network remains unchanged. Six conditions are evaluated: the full cross-layer model (CEDA), and five variants removing the network layer indicators, wind and physical hazard indicators, battery state, triage weights, and patient timer ratios independently.

\begin{figure*}[h!]
    \centering
    \includegraphics[width=1\linewidth]{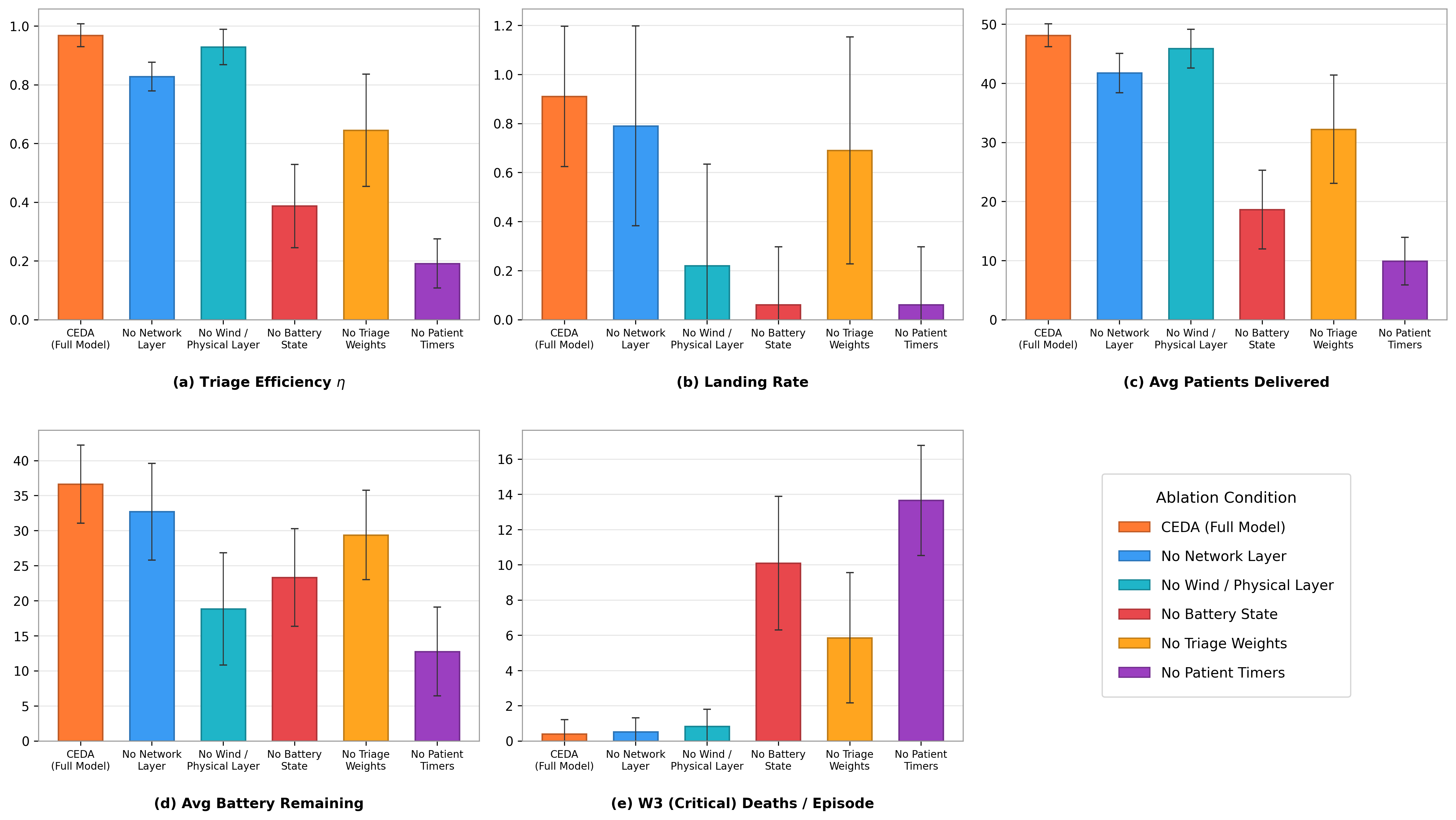}
    \caption{Cross-layer awareness ablation across five metrics. Each bar group shows the effect of removing one information layer from the agent's state vector at evaluation time while the policy remains unchanged. (a) Triage efficiency $\eta$. (b) Landing rate. (c) Average patients delivered per episode. (d) Average battery remaining. (e) W3 (Critical) deaths per episode.}
    \label{fig:awareness_ablation}
\end{figure*}

Figure~\ref{fig:awareness_ablation} reveals three tiers of information criticality. Removing network indicators produces only moderate degradation (triage efficiency falls approximately 14\%), reflecting a desirable deployment-time robustness in infrastructure-limited settings. Removing wind and hazard indicators preserves triage performance ($\eta = 0.93$) but collapses landing rate from 0.91 to 0.22 and nearly halves battery remaining, as the agent can no longer anticipate adverse conditions; removing triage weights instead leaves landing intact but reduces patients delivered from 48.1 to 32.2 and raises W3 unserved cases to 5.9, as Critical patients are no longer prioritized. Removing battery state or patient timers is most damaging: landing rate collapses to approximately 0.06 without battery state, while removing patient timers is the single worst ablation overall, with triage efficiency falling to 0.19 and W3 unserved cases reaching 13.7 per episode, suggesting the policy relies on deadline information for both sortie timing and patient selection.

Together, these results show that the physical layer governs mission completion and energy safety, the application scheduling layer governs service equity and urgency-sensitive prioritization (with patient timers the most consequential single input), and the network layer confers deployment-time robustness rather than a runtime sensor dependency; the full cross-layer state is necessary to simultaneously optimize all five metrics.

\section{PX4 Software-in-the-Loop Validation}
\label{sec:px4}

We evaluate whether CEDA's delivery, triage, hazard-avoidance, and recovery behavior persists when discrete actions are executed through the PX4 autopilot stack in software-in-the-loop (SITL) simulation. With five quadrotors, the experiments compare CEDA with heuristic dispatch and examine increased hazard density, reduced energy, and delayed action application. This is sim-to-sim validation: continuous simulated flight replaces idealized grid motion, while the task model and its transition safeguards remain in the execution loop.

\subsection{Experimental Setup}
\label{sec:px4_setup}

Five X500 quadrotors, each running an independent PX4 instance, operate in a shared headless Gazebo world through MAVSDK offboard position control. A coordinator updates the world model from live PX4 telemetry, queries the trained network, checks proposed grid transitions, and dispatches approved targets as 2\,m waypoint offsets in the north--east--down (NED) frame. Inference is orchestrated by this shared harness rather than independently deployed onboard; the target decision rate is 2\,Hz, extended by waypoint execution and telemetry handling, so decision steps are not fixed-duration physical intervals.

The harness retains the final-curriculum \texttt{environment} from Section~\ref{sec:results}: patient spawning and survival, hazards, energy accounting, landing eligibility, action masking, and rewards follow the training implementation, over a $100\times100$-cell grid, 50 patients, an 800-step mission budget, and a 400-step landing grace period. The world model rejects obstacle or out-of-bounds moves and reverts inter-drone conflicts before waypoint dispatch.

\textit{Nearest-Path} greedily assigns drone--patient pairs by ascending obstacle-aware path distance; \textit{Priority-Path} orders patients by triage weight, then remaining time, assigning each the nearest available drone; both use an energy-aware return rule based on the model available to CEDA. \textit{Random} samples uniformly from valid actions. In total, 200 SITL episodes were collected across nominal, hazard-density, delayed-action, and reduced-energy campaigns (full composition in Supplementary Table~S1). Policy comparisons use Welch's $t$-tests, with 95\% CIs computed as $\bar{x}\pm1.96s/\sqrt{n}$.

\subsection{Principal PX4 Comparison}
\label{sec:px4_headline}

Table~\ref{tab:px4_nominal} reports the nominal PX4 SITL comparison across all four policies. Relative to Nearest-Path, CEDA's delivery rate is higher by $+0.065$ ($p{=}0.0035$), its triage efficiency is higher by $+0.078$ ($p{=}0.0026$), and it recovers $+1.44$ more drones on average ($p{<}0.0001$); CEDA's wind and low-signal avoidance rates (0.998, 0.993) substantially exceed Nearest-Path's, with under one exposure step per episode. Priority-Path and Random perform substantially worse than either learned policy. In the abstract training environment, by contrast, neither the delivery-rate nor the triage-efficiency gap reaches significance, though CEDA already recovers more drones on average (4.96 vs.\ 3.92; full comparison in Supplementary Table~S2), so CEDA's advantage over Nearest-Path accompanies a substantial absolute delivery loss under PX4 execution, examined further below.

\begin{table*}[t]
\centering
\small
\caption{Nominal PX4 SITL outcomes, mean $\pm$ 95\% CI ($n{=}25$ CEDA and Nearest-Path; 10 Priority-Path and Random).}
\label{tab:px4_nominal}
\begin{tabular*}{0.94\textwidth}{@{\extracolsep{\fill}}lrrrr@{}}
\toprule
\textbf{Metric} & \textbf{CEDA} & \textbf{Nearest-Path} & \textbf{Priority-Path} & \textbf{Random} \\
\midrule
Delivery rate & $\mathbf{0.70} \pm 0.03$ & $0.64 \pm 0.03$ & $0.17 \pm 0.03$ & $0.01 \pm 0.01$ \\
Triage efficiency $\eta$ & $\mathbf{0.71} \pm 0.03$ & $0.64 \pm 0.04$ & $0.19 \pm 0.04$ & $0.01 \pm 0.01$ \\
Mortality rate & $\mathbf{0.30} \pm 0.03$ & $0.36 \pm 0.03$ & $0.83 \pm 0.03$ & $0.99 \pm 0.01$ \\
Drones recovered (of 5) & $\mathbf{4.28} \pm 0.37$ & $2.84 \pm 0.39$ & $2.90 \pm 0.80$ & $0.00$ \\
Drones depleted (of 5) & $\mathbf{0.72} \pm 0.37$ & $2.16 \pm 0.39$ & $2.10 \pm 0.80$ & $5.00$ \\
Wind avoidance & $\mathbf{0.998} \pm 0.002$ & $0.487 \pm 0.068$ & $0.553 \pm 0.059$ & $0.630 \pm 0.115$ \\
Low-signal avoidance & $\mathbf{0.993} \pm 0.011$ & $0.486 \pm 0.058$ & $0.497 \pm 0.053$ & $0.736 \pm 0.126$ \\
\bottomrule
\end{tabular*}
\end{table*}

\subsection{Transfer, Robustness, and Deployment Considerations}
\label{sec:px4_summary}

A matched-seed comparison between the abstract environment and PX4 SITL shows the expected cost of continuous-flight execution: episode length rises by approximately 40\%, delivery rate falls from 0.97 to 0.70, and recovered drones fall from 5.0 to 3.67 on average, while hazard-avoidance behavior does not deteriorate. Across a hazard-density sweep from $0.5\times$ to $2.0\times$ nominal density, CEDA's delivery-rate advantage over Nearest-Path grows from 3.0 to 12.0 percentage points, with wind avoidance (0.969--0.992) far above Nearest-Path's (0.517--0.535) throughout. At 60\% initial charge, CEDA retains a significant delivery-rate and triage-efficiency advantage over Priority-Path (0.338 vs.\ 0.078, $p{<}0.0001$), with no detected advantage in recovery. Sensitivity to delayed action is more severe: delivery rate falls from 0.702 at zero delay to 0.282 at a four-step delay, with invalid landing attempts rising sharply, a meaningful rather than negligible limitation. Full results, including execution diagnostics, matched-seed transfer, and the complete hazard-density, reduced-energy, and action-delay sweeps, are provided in Supplementary Sections~S4--S6.

The PX4 SITL campaign evaluates several, but not all, factors relevant to physical field deployment. Continuous flight dynamics are directly evaluated via the real PX4 autopilot stack against a Gazebo physics simulator, though this remains sim-to-sim validation, and onboard inference is not tested, as the trained network is queried by a shared external harness. Wind is modeled as discrete zones with elevated battery drain, but continuous aerodynamic disturbances are not modeled, and position feedback is treated as accurate with no localization error. Rather than modeling communication delay as a measured quantity such as MAVLink round-trip latency, we test the policy's sensitivity to its consequence directly, as above; the energy ledger deducts a fixed per-step drain independent of any carried payload. Weather uncertainty beyond discrete wind zones, GPS-derived localization error, and payload-dependent energy consumption remain open directions for future validation before physical deployment.

\section{Conclusion and Future Work}
\label{sec:conclusion}
In this paper, we address the compounding challenges of multi-drone medical supply delivery during disaster response via a unified cross-layered deep reinforcement learning framework that treats path planning, network awareness, and application scheduling as interdependent layers rather than independent subproblems.

We introduce CEDA, a Centralized Training with Decentralized Execution algorithm built on a Double Deep Q-Network augmented with a QMIX-style monotonic mixing network, jointly optimizing triage-priority-aware routing, multi-agent coordination, and energy-efficient navigation across all three layers. A Priority-Preserving Fair Scheduling strategy combines clinically defined triage classes, dynamic per-patient survival curves that escalate triage weight as a patient's condition deteriorates, timer-scaled delivery rewards, and outcome-level fairness and priority-ordering rewards, preserving priority ordering without starving lower-acuity patients. Cross-layer ablation experiments show no single information layer is solely responsible for CEDA's performance: the physical layer governs mission completion and energy safety, the application scheduling layer governs service equity and urgency-sensitive prioritization (with patient timer information the single most consequential input), and network-layer hazard indicators confer only modest deployment-time robustness, indicating the policy has largely internalized this information during training.

At an expanded scale of five drones and fifty patients, CEDA delivers over 95\% of spawned patients with a weighted triage efficiency exceeding 0.90 and a Jain's Fairness Index of approximately 1.0 across triage classes, while maintaining low and comparable mortality across all triage classes. Benchmarked against four contemporary MARL baselines (VDN, QMIX, MADDPG, MAPPO), CEDA leads the other reported task-performance metrics and matches MAPPO on agent collisions, with the strongest baseline, MAPPO, still trailing on delivery rate, triage efficiency, and mission success. PX4 software-in-the-loop validation with a five-vehicle fleet further confirms that CEDA's routing advantage remains executable through the PX4 autopilot stack: relative to a Nearest-Path heuristic baseline, CEDA achieves higher delivery rate and triage efficiency and recovers more of the fleet, while remaining sensitive to delayed action execution and reduced initial energy, both characterized directly rather than assumed away.

Future work will explore scaling to substantially larger deployments, replacing the fixed logistic survival curves with probabilistic deterioration forecasting for Critical patient prioritization under clinical uncertainty, and closing the sim-to-real gap through continuous aerodynamic wind modeling, GPS-derived localization error, payload-dependent energy consumption, delay-aware training, and continued outdoor PX4 validation with reduced simulation fidelity.


\balance
{\sloppy
\emergencystretch=20em
\bibliography{mybibfile}
}

\end{document}